\documentclass[letter]{article}

\usepackage[english]{babel}
\usepackage[utf8]{inputenc}
\usepackage[T1]{fontenc}

\usepackage[letterpaper,top=3cm,bottom=2cm,left=3cm,right=3cm,marginparwidth=1.75cm]{geometry}

\usepackage{amsmath}
\usepackage{graphicx}

\usepackage[colorlinks=true, allcolors=blue]{hyperref}
\usepackage{stackengine}
\usepackage{physics}
\usepackage{amssymb} 
\usepackage{mathrsfs}
\usepackage{amsfonts}
\usepackage{amssymb}
\usepackage[useregional]{datetime2}
\usepackage[left]{lineno}
\usepackage{longtable}
\usepackage{booktabs}
\usepackage[framemethod=TikZ]{mdframed}
\usepackage{tcolorbox}
\tcbuselibrary{breakable}
\usepackage{csquotes}
\usepackage{pdflscape}
\usepackage{tabularx}
\usepackage{longtable}
\usepackage{makecell}
\usepackage{adjustbox}
\usepackage{array}
\usepackage{csquotes}
\usepackage{blindtext}
\usepackage[affil-it]{authblk}

\usepackage[backend=biber, maxcitenames=2, uniquename=false, style=apa, sorting=nyt,doi=false,isbn=false,url=false, natbib=true]{biblatex}

\DeclareNameFormat{labelname:poss}{
  \nameparts{#1}
  \ifcase\value{uniquename}%
    \usebibmacro{name:family}{\namepartfamily}{\namepartgiven}{\namepartprefix}{\namepartsuffix}%
  \or
    \ifuseprefix
      {\usebibmacro{name:first-last}{\namepartfamily}{\namepartgiveni}{\namepartprefix}{\namepartsuffixi}}
      {\usebibmacro{name:first-last}{\namepartfamily}{\namepartgiveni}{\namepartprefixi}{\namepartsuffixi}}%
  \or
    \usebibmacro{name:first-last}{\namepartfamily}{\namepartgiven}{\namepartprefix}{\namepartsuffix}%
  \fi
  \usebibmacro{name:andothers}%
  \ifnumequal{\value{listcount}}{\value{liststop}}{'s}{}}
\DeclareFieldFormat{shorthand:poss}{%
  \ifnameundef{labelname}{#1's}{#1}}
\DeclareFieldFormat{citetitle:poss}{\mkbibemph{#1}'s}
\DeclareFieldFormat{label:poss}{#1's}
\newrobustcmd*{\posscitealias}{%
  \AtNextCite{%
    \DeclareNameAlias{labelname}{labelname:poss}%
    \DeclareFieldAlias{shorthand}{shorthand:poss}%
    \DeclareFieldAlias{citetitle}{citetitle:poss}%
    \DeclareFieldAlias{label}{label:poss}}}
\newrobustcmd*{\posscite}{%
  \posscitealias%
  \textcite}
\newrobustcmd*{\Posscite}{\bibsentence\posscite}
\newrobustcmd*{\posscites}{%
  \posscitealias%
  \textcites}

\newcommand{\E}[2]{\mathbb{E}_{#1} \negmedspace \left[ #2 \right]}
\newcommand{\Cov}[3]{\mathrm{Cov}_{#1} \negmedspace \left( #2, #3 \right)}
\newcommand{\Var}[2]{\mathrm{Var}_{#1} \negmedspace \left( #2\right)}
\renewcommand{\eqref}[1]{Eq.\ref{#1}}

\title{Coexistence in spatiotemporally fluctuating environments}
\date{\today}

\author[1,2,*]{Evan C. Johnson}
\author[1]{Alan Hastings}

\affil[1,]{Department of Environmental Science and Policy; University of California Davis; Davis, California 95616 USA}
\affil[2,]{Center for Population Biology; University of California Davis; Davis, California 95616 USA}
\affil[*]{Corresponding author: Evan Johnson, evcjohnson@ucdavis.edu}

\begin{document}
\maketitle
\clearpage

\section*{Abstract}

Ecologists have put forward many explanations for coexistence, but these are only \textit{partial explanations}; nature is complex, so it is reasonable to assume that in any given ecological community, multiple mechanisms of coexistence are operating at the same time. Here, we present a methodology for quantifying the relative importance of different explanations for coexistence, based on an extension of \textit{Modern Coexistence Theory}. Current versions of Modern Coexistence Theory only allow for the analysis of communities that are affected by spatial \textit{or} temporal environmental variation, but not both. We show how to analyze communities with spatiotemporal fluctuations, how to parse the importance of spatial variation and temporal variation, and how to measure everything with either mathematical expressions or simulation experiments. Our extension of Modern Coexistence Theory allows empiricists to use realistic models and more data to better infer the mechanisms of coexistence in real communities. 
\linebreak
\linebreak
Keywords: modern coexistence theory,  spatiotemporal, environmental variation, environmental stochasticity, storage effect, relative nonlinearity, fitness-density covariance.

\tableofcontents

\newpage

\begin{longtable}[t]{p{0.3\linewidth}  p{0.7\linewidth} }
\caption{The symbols and terminology of Spatiotemporal Modern Coexistence Theory (MCT).} \\
\toprule \\
 & \multicolumn{1}{c}{Description}  \\
\hline \\
\multicolumn{2}{l}{MCT-specific terminology} \\ 
\cmidrule(lr){1-2} \\
invader & a rare species; for mathematical convenience, the per capita growth rate of this species is approximated by perturbing population density to zero.  \\
resident & a common species, more precisely understood as a species at its typical abundances \\
invasion growth rate & the long-term average of the per capita growth rate of an invader \\
partition & a scheme for breaking up an invasion growth rate into a sum of component parts \\
coexistence mechanism & a class of explanations for coexistence; corresponds to a component of the invasion growth rate partition of Spatiotemporal MCT \\
space-time decomposition & a type of partition which attempts to parse the effects of spatial and temporal variation on the invasion growth rate \\
invader--resident comparison & a comparison between an invader and the resident species; measures a rare-species advantage  \\
speed conversion factor & converts the population-dynamical speed of resident species to that of the invader; corrects for average fitness differences in the invader--resident comparison; replaces the \textit{scaling factors}, also known as \textit{comparison quotients}, from previous versions of MCT  \\
\end{longtable}
\begin{longtable}[t]{p{0.1\linewidth}  p{0.8\linewidth}}
\midrule \\ 
\multicolumn{2}{l}{Variable} \\ 
\cmidrule(lr){1-2} \\
$x$ & a location in space \\
$t$ & a point in time \\
$j$ & species index (subscript) \\
$n_j(x,t)$ & the population density of species $j$ at patch $x$ and time $t$. \\
$\nu_j(x,t)$ & relative density, calculated as local population density divided by the spatial average of population density, i.e., $n_j(x,t) / \E{x}{n_j}$ \\
$\lambda_j$(x,t) & the local finite rate of increase. In non-spatial models, $\lambda_j$ is defined as $n_{j}(x,t+1)/n_{j}(x,t)$. However, in spatial models, $\lambda_j$ is defined as $n_{j}'(x,t)/n_{j}(x,t)$, where $n_{j}'(x,t)$ is the population size after the local growth phase, but before the dispersal phase. \\
$\widetilde{\lambda}_j(t)$ & the metapopulation finite rate of increase, defined as a density-weighted average over patches: $\widetilde{\lambda}_j = \E{x}{(n_j / \E{x}{n_j}) \lambda_j}$ \\
$\E{t}{\log(\widetilde{\lambda}_j)}$ & The long-term average growth rate; for resident species, this is zero by definition; for invader, this is the invasion growth rate \\
$E_j(x,t)$ & the environmental parameter; more generally understood as the effects of density-independent factors \\
$C_j(x,t)$ & the competition parameter; more generally understood as the effects of density-dependent factors \\
$g_j$ & a function that gives the local finite rate of increase: $\lambda_j(x,t) = g_j(E_j(x,t), C_j(x,t))$ \\
$E_j^*$ & the equilibrium environmental parameter, defined so that $g_j(E_j^*, C_j^*) = 1$ \\
$C_j^*$ & the equilibrium competition parameter, defined so that $g_j(E_j^*, C_j^*) = 1$ \\
$\sigma$ & the scale of environmental fluctuations: $E_j(x,t) - E_j^* = \mathcal{O}(\sigma)$; it is often the case that $\sigma$ controls the size of fluctuations in $n_j$, $E_j$, and $C_j$, see Appendix \ref{app:Deriving small-noise coexistence mechanisms:Small noise assumptions} \\
$S$ & the total number of species in the community; $S-1$ is the number of residents \\
$a_j$ & the \textit{speed} of the population dynamics of species $j$; the intrinsic capacity to grow or decline quickly; often operationalized $a_j = 1/GT_j$, where $GT_j$ is the generation time of species $j$  \\
$\frac{a_i}{a_j}$ & speed conversion factors; a constant that effectively converts the population-dynamical speed of species $j$ to that of species $i$ \\
$\overline{\mathscr{E}_j}'$ & the main effect of density-independent factors on the invasion growth rate, defined as $\E{t}{\log(\E{x}{g_j(E_j,C_j^*)})}$  \\
$\overline{\mathscr{C}_j}'$ & the main effect of density-dependent on the invasion growth rate, defined as $\E{t}{\log(\E{x}{g_j(E_j^*, C_j)})}$  \\ 
$\overline{\mathscr{I}_j}'$ & the interaction effect density-dependent and density-independent factors on the invasion growth rate, defined as $\E{t}{\log(\E{x}{g_j})} - \overline{\mathscr{E}_j}' - \overline{\mathscr{C}_j}'$  \\
\midrule \\
\multicolumn{2}{l}{Coexistence mechanisms} \\
\cmidrule(lr){1-2} \\
$\Delta E_i$ & Density-independent effects; the degree to which density-independent factors favor the invader \\
$\Delta \rho_i$ & Linear density-dependent effects; specialization on resources and/or natural enemies\\ 
$\Delta N_i$ & Relative nonlinearity; specialization on the spatiotemporal variance of resources and/or natural enemies  \\ 
$\Delta I_i$ & The storage effect; specialization on different states of a spatiotemporally varying environment  \\ 
$\Delta \kappa_i$ & Fitness-density covariance; the differential ability of rare species to end up in locations with high ecological fitness  \\
\midrule \\
\multicolumn{2}{l}{Taylor series coefficients} \\ 
\cmidrule(lr){1-2} \\
$\alpha_{j}^{(1)}$ & the linear effects of fluctuations in $E_j$, defined as $\pdv{g_j}{E_j}\Bigr|_{\substack{E_j = E_{j}^{*}\\C_j = C_{j}^{*}}} = \pdv{g_j\scriptstyle{(E_j^*, C_j^*)}}{E_j}$ \\ 
$\alpha_{j}^{(2)}$ & the nonlinear effects of  of fluctuations in $E_j$, defined as $\pdv[2]{g_j\scriptstyle{(E_j^*, C_j^*)}}{E_j}$  \\ 
$\beta_{j}^{(1)}$ & the linear effects of fluctuations in $C_j$, defined as $\pdv{g_j\scriptstyle{(E_j^*, C_j^*)}}{C_j}$  \\ 
$\beta_{j}^{(2)}$ & the nonlinear effects of fluctuations in $C_j$, defined as $\pdv[2]{g_j\scriptstyle{(E_j^*, C_j^*)}}{C_j}$  \\ 
$\zeta_{j}^{(1)}$ & the non-additive (i.e., interaction) effects of fluctuations in $E_j$ and $C_j$, defined as $\zeta_j = \pdv{g_j\scriptstyle{(E_j^*, C_j^*)}}{E_j}{C_j}$  \\ 
\midrule \\ 
\multicolumn{2}{l}{Superscripts and subscripts} \\ 
\cmidrule(lr){1-2} \\
Subscripts &  \\ 
\cmidrule(lr){1-1} \\
$j$ & index of an arbitrary species   \\
$i$  & index of the invader  \\
$r$ & index of a resident  \\
$x$ & indicates that a summary statistic (e.g., mean, covariance, variance) is calculated by summing across space  \\
$t$ & indicates that a summary statistic (e.g., mean, covariance, variance) is calculated by summing across time \\
$A$ & denotes the effect of average conditions in the space-time decomposition  \\ 
$S$ & denotes the main effect of spatial variation in the space-time decomposition  \\
$T$ & denotes the main effect of temporal variation in the space-time decomposition  \\
$R$ & denotes the interaction effect of spatial and temporal variation in the space-time decomposition  \\
Superscripts &  \\ 
\cmidrule(lr){1-1} \\
$(e)$ & denotes exact coexistence mechanisms, or intermediate products in the calculation of exact coexistence mechanisms  \\
$\#$ & indicates that the elements of a vector or matrix have been randomized (sampled randomly without replacement)  \\
\midrule  \\ 
\multicolumn{2}{l}{Operators} \\ 
\cmidrule(lr){1-2} \\
$\E{x,t}{\cdot}$ & The spatiotemporal \textit{sample} arithmetic mean; for a variable $Z$ that varies over $K$ patches and $T$ time points, \\ 
& $\E{x}{Z} = (1/K)\sum_{x = 1}^{K} Z(x,t)$, \\
& $\E{t}{Z} = (1/T)\sum_{t = 1}^{T} Z(x,t)$, and \\
& $\E{x,t}{Z} = (1/(T K)) \sum_{t = 1}^{T} \sum_{x = 1}^{K} Z(x,t)$  \\
$\Var{x,t}{\cdot}$ & The spatiotemporal \textit{sample} variance for a variable $Z$ that varies over $K$ patches and $T$ time points, \\
& $\Var{x}{Z} = (1/K)\sum_{x = 1}^{K} (Z(x,t) - \E{x}{Z})^2$, \\
& $\Var{t}{Z} = (1/T)\sum_{t = 1}^{T} (Z(x,t) - \E{t}{Z})^2$, and \\
& $\Var{x,t}{Z} = (1/(T K)) \sum_{t = 1}^{T} \sum_{x = 1}^{K} (Z(x,t) - \E{x,t}{Z})^2$ \\
$\Cov{x,t}{\cdot}{\cdot}$ & The spatiotemporal \textit{sample} covariance of variables $W$ and $Z$ that vary over $K$ patches and $T$ time points,  \\
& $\Cov{x}{W}{Z} = (1/K)\sum_{x = 1}^{K} (W(x,t) - \E{x}{W})(Z(x,t) - \E{x}{Z})$, \\
& $\Cov{t}{W}{Z} = (1/T)\sum_{t = 1}^{T} (W(x,t) - \E{t}{W})(Z(x,t) - \E{t}{Z})$, and \\
& $\Cov{x,t}{W}{Z} = (1/(T K)) \sum_{t = 1}^{T} \sum_{x = 1}^{K} (W(x,t) - \E{x,t}{W})(Z(x,t) - \E{x,t}{Z})$ \\
\bottomrule \\ 
\end{longtable}

\section{Introduction}

Modern Coexistence Theory (MCT) is a framework for understanding ecological coexistence (\cite{Chesson1994}; \cite{chesson2000general}; see \cite{barabas2018chesson} for a recent review). MCT has two main strengths. First, MCT gives us the relative importance of different explanations for coexistence, and thus tells us \textit{how} species are coexisting (not simply whether they \textit{are} coexisting). Second, MCT is \textit{general} because it is framework for analyzing arbitrary models of population dynamics (which could represent all kinds of different communities). This feature of MCT stands in contrast to several big theories in community ecology --- such as neutral theory, maximum entropy, and metacommunity theory --- in which a small number of models are used to make inferences about many communities. MCT has been successfully used to derive theoretical insights (e.g., \cite{chesson1997roles}; \cite{stump2015distance}; \cite{li2016effects}; \cite{snyder2003local}; \cite{Chesson2008}; \cite{kuang2010interacting}; \cite{schreiber2021positively}), and has been used to determined the mechanisms of coexistence in real communities (\cite{Caceres1997}; \cite{Adler2006}; \cite{Angert2009}; \cite{Sears2007}; \cite{usinowicz2012coexistence}; \cite{descamps2005stable}; \cite{chu2015large}; \cite{Usinowicz2017}; \cite{ignace2018role}; \cite{towers2020requirements}).

Despite MCT's successes, there are a handful of problems that limit its applicability. One such problem is that currently, MCT can be used analyze models where the environment fluctuates over space or time, but not both. \textbf{Here, we extend Modern Coexistence Theory (MCT) to show how models with spatiotemporal fluctuations can be analyzed. Further, we show how to parse the importance of spatial fluctuations and temporal fluctuations, and how to measure everything with mathematics or simulations}. While a couple papers (\cite{Chesson1985}; \cite{snyder2005examining}; \cite{Snyder2008}) have examined the effects of spatiotemporal fluctuations in particular models, our approach permits the analysis of a broad variety of models and is thus targeted towards empirical applications.  

The ability to analyze models with spatiotemporal fluctuations can lead to novel theoretical insights. For instance, we show that 1) temporal variation can promote the storage effect in the lottery model, even in the case of non-overlapping generations (Section \ref{sec:Example: the spatiotemporal lottery model}), 2) that it is (nearly) impossible for the competitive exclusion principle to hold true in the presence of spatiotemporal fluctuations (Section \ref{sec:Discussion}), and 3) the inclusion of spatiotemporal fluctuations exactly doubles the maximum number of species that can coexist due to fluctuation-dependent coexistence mechanisms (Section \ref{sec:Discussion}). 

More importantly, the ability to analyze models with spatiotemporal fluctuations helps us better understand coexistence in real ecological communities: MCT necessarily interfaces with the real world through empirically-calibrated models, and good representations of real communities will undoubtedly involve spatiotemporal variation. But the addition of spatiotemporal fluctuations is not realism for realism's sake: failure to include spatiotemporal fluctuations will typically lead to underestimates of fluctuation-dependent coexistence mechanisms, which could lead to poor downstream inferences about the nature of coexistence, community composition, and species abundance distributions. Our extension of MCT will enable ecologists to use more data and better models to understand coexistence.

\section{Spatiotemporal coexistence mechanisms}
\label{sec:Spatiotemporal coexistence mechanisms}

\subsection{Overview}
\label{sec:Spatiotemporal coexistence mechanisms:Overview}

At the coarsest level of description, Modern Coexistence Theory (MCT) has two steps: "decompose and compare" (\cite{Ellner2019}). First, \textit{decompose} the long-term average per capita growth rate of each species into terms that correspond to conceptually distinct processes (e.g., growth that can be attributed to resource consumption).  Second, \textit{compare} the like-terms of rare species (termed \textit{invaders}) and common species (termed \textit{residents}) in order to discover which processes tend to help rare species. These invader--resident comparisons, called \textit{coexistence mechanisms}, correspond to classes of explanations for coexistence. The sum of coexistence mechanisms is the \textit{invasion growth rate}, the long-term average per capita growth rate of a species that has been perturbed to near-zero density.

How do invasion growth rates and coexistence mechanisms relate to coexistence? The main idea is that invasion growth rates measure the tendency to recover from rarity, so a set of $S$ species can be said to coexist if each species has a positive invasion growth rate in the sub-community of $S-1$ resident species. This is known as the \textit{mutual invasibility criterion} for coexistence (\cite{turelli1978reexamination}; \cite{chesson2000general}; \cite{Chesson1989}; \cite{Grainger2019TheResearch}). 

In truth, the relationship between invasion growth rates and coexistence is not so simple. The mutual invasibility criterion fails when the elimination of one species causes knock-on extinctions, such that the $S-1$ residents cannot coexist. For the mutual invasibility criterion to work, we must either assume that all $S-1$ residents can coexist (\cite{Case2000}), or limit ourselves to two-species competitive communities (\cite{Ellner1989}). When the mutual invasibility criterion fails, one can still use invasion growth rates as inputs to the Hofbauer criterion for coexistence (\cite{hofbauer1981general}; \cite{Benaim2019}, Eq.3.4), a sufficient condition for a type of global stability called \textit{permanence} or \textit{uniform persistence} (\cite{Schreiber2000}; \cite{garay2003robust}; \cite{Schreiber2011}; \cite{Roth2014}). But this criterion potentially combines invasion growth rates in many sub-communities (with $S-n$ residents, $n = 1, 2, \ldots, S$), so it is unclear to how to average over sub-communities to obtain species-level coexistence mechanisms or community-average coexistence mechanisms (as in \cite{Chesson2003}, Eq.16).

Invasion growth rates are used in the mutual invasibility criterion and the Hofbauer criterion, both of which test for \textit{global stability}. However, global stability can sometimes be too strong a notion of coexistence: under a certain set of scenarios (e.g., allee effects, obligate mutualisms, and intransitive competition) negative invasion growth rates can erroneously indicate a failure to coexistence, since all species would be able to coexist if simultaneously introduced at higher densities. We leave all of these issues to future research; thus, we temporarily use these concepts heuristically: larger coexistence mechanisms $\rightarrow$ larger invasion growth rate $\rightarrow$ stronger coexistence. 

In the MCT literature, there are two types of \textit{coexistence mechanisms}. The first is \textit{small-noise coexistence mechanisms}, which closely approximate the invasion growth rate when environmental fluctuations are small. The second type is \textit{exact coexistence mechanisms}, which always sum exactly to the invasion growth rate. Both types of coexistence mechanisms have been used in previous papers, though they have not yet been explicitly named or differentiated; in fact, most expositions of MCT present partitions of the invasion growth rate that mix-and-match both types of coexistence mechanisms (e.g., \cite{barabas2018chesson}, Eq.19; \cite{Chesson1994}, Eq. 19--22).
To be clear, small-noise coexistence mechanisms do no assume that environmental fluctuations are unimportant or that the fluctuation-independent mechanisms drive coexistence. \textit{Small-noise} refers to a technical assumption that environmental fluctuations are small relative to other parameters in a model of population growth. This assumption, (when paired some additional assumptions; Appendix \ref{app:Deriving small-noise coexistence mechanisms:Small noise assumptions}) allows us to derive analytical expressions for the coexistence mechanisms.

Even though small-noise coexistence mechanisms only approximate the invasion growth rate, there are situations in which small-noise coexistence mechanisms are preferred. For one, the small-noise approximations can be calculated quickly, which is important in empirical applications where coexistence mechanisms are calculated for many draws from a posterior or bootstrap distribution of model parameters. Secondly, small-noise coexistence mechanisms sometimes permit analytical expressions (for a worked example, see Section \ref{sec:Example: the spatiotemporal lottery model}), whereas the exact coexistence mechanisms almost never do. Finally, the small-noise coexistence mechanisms could correspond more closely to our verbal/textual explanations for coexistence, and thus could be more interpretable. On the other hand, the primary boon of the exact coexistence mechanisms is that they sum exactly to the invasion growth rate. We will derive both the small-noise coexistence mechanism (Section \ref{sec:Spatiotemporal coexistence mechanisms:Small-noise coexistence mechanisms}) and the exact coexistence mechanisms (Section \ref{sec:Spatiotemporal coexistence mechanisms:Exact coexistence mechanisms}), but we leave it to the reader to determine which is more relevant to their work. 

Throughout this paper we will consider population with spatial structure but without age/stage structure, whose dynamics operate in discrete time. In Appendix \ref{app:Generalization of MCT to different classes of models}, we discuss generalizations of Spatiotemporal MCT to different classes of models, including continuous-time models and age/stage-structured population models. For the time being, community dynamics are governed by a system of difference equations, 
\begin{equation} \label{local_lambda}
n_j(x,t+1) = n_{j}(x,t) \;  \lambda_{j}(x,t) + m_j(x,t) - e_j(x,t) \qquad j = (1, 2, ..., S),
\end{equation}
where $n_{j}(x,t)$ is the local density of species $j$, $\lambda_j$ is the local finite rate of increase of species $j$, $x$ is a discrete patch in space, $t$ is a discrete point in time, and $S$ is the number of species in the community. The local finite rate of increase is a function of the effects of the environment $E_j$ and competition $C_j$, which themselves may fluctuate over space and time. The terms $m_j$ and $e_j$ represent immigration and emmigration respectively, in units of population density. We require that the sum of $c_j$ and $e_j$ across space (i.e., net dispersal) vanishes (Appendix \ref{app:Deriving small-noise coexistence mechanisms:Spatial averaging and fitness-density covariance}), which occurs generically when either 1) the system is \textit{closed} (i.e., no individuals can enter or leave the system of patches), or 2) that the system of patches is representative or a larger metacommunity, such that it receives roughly as many immigrants as it loses emigrants.

A few notes on notation are necessary. For convenience, we will often write out random variables without the explicit dependence on space and time; for example, we will write $\lambda_j$ instead of $\lambda_j(x,t)$. We use the operator $\E{}{Z}$ to denote the averaging (i.e., the sample mean) of some random variable $Z$, with a subscript to denote whether the average is being taken across space, time, or both. For example, in a system with $K$ patches that has been observed for $T$ time-steps, $\E{x}{Z} = (1/K)\sum_{x = 1}^{K} Z(x,t)$, $\E{t}{Z} = (1/T)\sum_{t = 1}^{T} Z(x,t)$, and $\E{x,t}{Z} = (1/(T K)) \sum_{t = 1}^{T} \sum_{x = 1}^{K} Z(x,t)$. This notation is unorthodox for two reasons. First, the spatial dependence of temporal average, $(1/T)\sum_{t = 1}^{T} Z(x,t)$, is suppressed by the notation $\E{t}{Z}$ (a similar thing can be said of the spatial average). Second, the expectation operator conventionally denotes the average across an infinite number of instantiations of the stochastic population process at one point in time; not the temporal average of one instantiation (though they are asymptotically equivalent if the stochastic process is stationary and ergodic; see Section \ref{sec:Extended discussion of invasion growth rates, including computational tricks}). We define $\Var{}{.}$ and $\Cov{}{.}{.}$ in a similar fashion, to the denote the sample variance and sample covariance respectively. 

The \textit{local finite rate of increase}, $\lambda_j$, is the discrete-time analogue of a per capita growth rate. The \textit{metapopulation finite rate of increase}, $\widetilde{\lambda}_j = \E{x}{(n_j / \E{x}{n_j}) \lambda_j}$, is the density-weighted average of $\lambda_j$ across space. The \textit{average growth rate} rate, $\E{t}{\log(\widetilde{\lambda}_j)}$, is the quantity which is predictive of long-term growth (\cite{Lewontin1969}; \cite{Dempster1955}; \cite{Stearns2000}). The average growth rate of the invader species is called the \textit{invasion growth rate}. The subscript $i$ references an invader species, the subscript $r$ references a resident species, and the subscript $j$ references a generic species whose status as a resident or invader is impertinent.

\subsection{Small-noise coexistence mechanisms}
\label{sec:Spatiotemporal coexistence mechanisms:Small-noise coexistence mechanisms}

A full derivation of small-noise spatiotemporal coexistence mechanisms is provided in Appendix \ref{app:Deriving small-noise coexistence mechanisms}. Here, we summarize the main steps:

\begin{enumerate}
    
    \item The local finite rate of increase is expressed as a function of an environmental parameter $E_j$, and a competition parameter $C_j$: $\lambda_j(x,t) = g_j(E_j(x,t), C_j(x,t))$.
    
   In past literature, the environmental parameter $E_j$ is also known as "the environmentally-dependent parameter", "the environmental response", or simply, "the environment". It is more generally defined as some parameter that depends on spatiotemporally fluctuating density-independent factors (e.g., the germination probability of a seed, which depends on precipitation). Similarly, the competition parameter $C_j$, also known as "competition", is more generally defined as some parameter that depends on density-dependent factors. As such, the competition parameter may resource competition, apparent competition, or even mutualism. The competition parameter can often be expressed as function of multiple regulating factors, such as resources, refugia, competitors' densities (as in the Lotka-Volterra model), and predators (see Appendix \ref{app:Generalization of MCT to different classes of models:Continuous-time models:Multiple regulating factors}).

    \item  The local finite rate of increase is approximated with a second-order Taylor series expansion of $g_j$ about the \textit{equilibrium parameters}, $E_j^*$ and $C_j^*$, which are specified by the user of MCT but must satisfy the constraint $g_j(E_j^*, C_j^*) = 1$. The resulting second-order polynomial will lead to an accurate approximation of the invasion growth rate, but only if some assumptions about the magnitude of environmental fluctuations are met (see Appendix \ref{app:Deriving small-noise coexistence mechanisms:Small noise assumptions}). To help satisfy these assumptions, it is important to select the equilibrium parameters so that they are close to their spatiotemporal means, $\E{x,t}{E_j}$ and $\E{x,t}{C_j}$ respectively.
    
    \item The appropriate spatial and temporal averaging is applied in order to express average growth rates entirely in terms of moments of local growth, $\lambda_j$, and relative density, $\nu_j = n_j / \E{x}{n_j}$:

\begin{equation} \label{time_decomp_2}
\begin{aligned}
   \E{t}{\log(\widetilde{\lambda}_j)} \approx \E{x,t}{\lambda_j} + \E{t}{\Cov{x}{\nu}{\lambda_j}} - 1 -   \frac{1}{2} \Var{t}{\E{t}{\lambda_j}}
\end{aligned}
\end{equation}

    \item The Taylor series approximation of $\lambda_j$ (see step 1) is substituted into the expression for the average growth rate (\eqref{time_decomp_2}), resulting in a long expression for species $j$'s average growth rate:
    
    \begin{equation} \label{big_decomp}
\begin{aligned}
\left. {\E{t}{\log(\widetilde{\lambda}_j)}}%
_{\stackunder[1pt]{}{}}%
\right|_{%
\stackon[1pt]{$\scriptscriptstyle E_j = E_{j}^{*}$}{$\scriptscriptstyle C_j = C_{j}^{*}$}}
\approx \; &  \alpha_j^{(1)} \E{x,t}{(E_j - E_{j}^{*})} + \beta_j^{(1)} \E{x,t}{(C_j - C_{j}^{*})} \\ 
+ & \; \frac{1}{2} \alpha_j^{(2)} \Var{x,t}{E_j} + \frac{1}{2} \beta_j^{(2)} \Var{x,t}{C_j} + \zeta_j  \Cov{x,t}{E_j}{C_j} \\ 
+ & \; \E{t}{\Cov{x}{\nu_j}{ \alpha_j^{(1)} (E_j - E_{j}^{*}) + \beta_j^{(1)} (C_j - C_{j}^{*})}} \\ 
- & \; \frac{1}{2} \alpha_j^{(1)^{2}} \Var{t}{\E{x}{E_j}} - \frac{1}{2} \beta_j^{(1)^{2}} \Var{t}{\E{x}{E_j}} - \alpha_j^{(1)}\beta_j^{(1)} \Cov{t}{\E{x}{E_j}}{\E{x}{C_j}},
\end{aligned}
\end{equation}

where the coefficients of the Taylor series, 

\begin{equation}  \label{taylor_coef}
\begin{aligned}
 \alpha_j^{(1)} = \pdv{g_j\scriptstyle{(E_j^*, C_j^*)}}{E_j},  \quad
 \beta_j^{(1)} = \pdv{g_j\scriptstyle{(E_j^*, C_j^*)}}{C_j},  \quad
 \alpha_j^{(2)} = \pdv[2]{g_j\scriptstyle{(E_j^*, C_j^*)}}{E_j},  \quad
 \beta_j^{(2)} = \pdv[2]{g_j\scriptstyle{(E_j^*, C_j^*)}}{C_j,},  \quad
 \zeta_j = \pdv{g_j\scriptstyle{(E_j^*, C_j^*)}}{E_j}{C_j},  \quad
\end{aligned}
\end{equation}

are all evaluated at user-specified equilibrium values $E_j = E_j^*$ and $C_j = C_j^*$, as implied by the notation.

The additive terms in the above equation (\eqref{big_decomp}), which we may call \textit{growth rate components}, can be conceptualized as distinct processes. For example, the second term $\beta_j^{(1)} \E{x,t}{(C_j - C_{j}^{*})}$ is the effect of the mean level of competition on the average growth rate.

\item The invader is compared to the residents. Because coexistence is about a \textit{rare-species advantage}, we do not care so much about the invader's growth rate components, but rather their magnitude relative to the corresponding components of residents. Since every resident species cannot grow or decline on average (i.e., $\E{t}{\log(\widetilde{\lambda}_r)} = 0$) we may subtract a linear combination of the $S-1$ resident species from the invasion growth rate

\begin{equation} \label{inv_res_dif}
    \E{t}{\log(\widetilde{\lambda}_i)} = \E{t}{\log(\widetilde{\lambda}_i)} - \frac{1}{S-1} \sum\limits_{r \neq i}^S \frac{a_i}{a_r} \E{t}{\log(\widetilde{\lambda}_r)},
\end{equation}

without any distortion of the invasion growth rate. The coefficients $a_i/a_r$ are called \textit{speed conversion factors} (\cite{johnson2022methods}) and virtually covert the population-dynamical speed of the residents to that of the invader. They will be discussed further in a few paragraphs. The long decomposition of the average growth rate (\eqref{big_decomp}) can be substituted into the above equation, and like-terms can be grouped such that the invasion growth rate is expressed as a sum of invader--resident comparisons. These comparisons are the \textit{coexistence mechanisms}. 
\end{enumerate}

\begin{tcolorbox}[breakable, title = Formulas for small-noise coexistence mechanisms, subtitle style={boxrule=0.4pt, colback=black!30!white}]

\tcbsubtitle{The invasion growth rate}

\begin{flalign} \label{sn_co_mech}
    \E{t}{\log(\widetilde{\lambda}_{i})} \approx \Delta E_{i} + \Delta\rho_i + \Delta\mathrm{N}_i + \Delta\mathrm{I}_i + \Delta\kappa_i
\end{flalign}

\tcbsubtitle{Density independent effects}

\begin{flalign}
    \label{dE}
    \Delta E_{i} = & \left[\alpha_{i}^{(1)} \E{x,t}{E_i - E_{i}^{*}} + \frac{1}{2} \alpha_i^{(2)}\Var{x,t}{E_i} - \frac{1}{2} \alpha_i^{(1)^{2}} \Var{t}{\E{x}{E_i}}\right] \nonumber \\ 
     & - \sum\limits_{r \neq i}^S \frac{a_i}{a_r} \left[\alpha_r^{(1)} \E{x,t}{E_r - \frac{1}{S-1} E_{r}^{*}} + \frac{1}{2} \alpha_r^{(2)}\Var{x,t}{E_r} - \frac{1}{2} \alpha_r^{(1)^{2}} \Var{t}{\E{x}{E_r}}\right] 
\end{flalign}

\tcbsubtitle{Linear density-dependent effects}
      
\begin{flalign} \label{drho}
   \Delta\rho_i = & \beta_i^{(1)} \E{x,t}{C_i - C_{i}^{*}}  - \frac{1}{S-1} \sum\limits_{r \neq i}^S \frac{a_i}{a_r} \beta_r^{(1)} \E{x,t}{C_r - C_{r}^{*}} 
\end{flalign}

\tcbsubtitle{Relative nonlinearity}
      
\begin{flalign} \label{dN}
     \Delta\mathrm{N}_i = & \frac{1}{2} \left[ \beta_i^{(2)}\Var{x,t}{C_i} - \beta_i^{(1)^{2}} \Var{t}{\E{x}{C_i}}\right] \nonumber  \\
    & - \frac{1}{S-1} \sum\limits_{r \neq i}^S \frac{a_i}{a_r} \left[ \beta_r^{(2)}\Var{x,t}{C_r} - \beta_r^{(1)^{2}} \Var{t}{\E{x}{C_r}}\right]
\end{flalign}

\tcbsubtitle{The storage effect}
      
\begin{flalign} \label{dI}
     \Delta\mathrm{I}_i =  & \left[ \zeta_i \Cov{x,t}{E_i}{C_i} - \alpha_i^{(1)}\beta_i^{(1)} \Cov{t}{\E{x}{E_i}}{\E{x}{C_i}}\right] \nonumber  \\
   & - \frac{1}{S-1} \sum\limits_{r \neq i}^S \frac{a_i}{a_r} \left[ \zeta_r \Cov{x,t}{E_r}{C_r} - \alpha_r^{(1)}\beta_r^{(1)} \Cov{t}{\E{x}{E_r}}{\E{x}{C_r}}\right]
\end{flalign} 

\tcbsubtitle{Fitness-density covariance}
      
\begin{flalign} \label{dkappa}
    \Delta\kappa_i = & \E{t}{\Cov{x}{\nu_i}{ \alpha_i^{(1)} E_i + \beta_i^{(1)} C_i}} \nonumber  \\
    & - \frac{1}{S-1} \sum\limits_{r \neq i}^S \frac{a_i}{a_r} \E{t}{\Cov{x}{\nu_r}{ \alpha_r^{(1)} E_r + \beta_r^{(1)} C_r}}
\end{flalign}     
\end{tcolorbox}

The interpretations of the coexistence mechanisms are as follows. Density independent effects ($\Delta E_i$) is the degree to which all density-independent factors favor the invader. The linear density-dependent effects ($\Delta \rho_i$) represents a rare-species advantage due to specialization on regulating factors (i.e., resources and/or natural enemies). Relative nonlinearity ($\Delta N_i$) is a rare-species advantage due to specialization on variation in regulating factors. The storage effect is the rare-species advantage due to specialization on certain states of a variable environment. Fitness-density covariance ($\Delta \kappa$) is the differential ability of a rare species’ individuals to end up in locations where they have high fitness. Note that "coexistence mechanism" is a misnomer when it comes to $\Delta E_i$, since $\Delta E_i$ can only support a single species in the absence of all other mechanisms. See \textcite{barabas2018chesson} for a more thorough discussion of the coexistence mechanisms and their interpretations.

Experts in coexistence theory may notice two differences between spatiotemporal MCT and previous versions of MCT (i.e., \cite{Chesson1994}, \cite{chesson2000general}), aside from the inclusion of spatiotemporal fluctuations. First, we keep the equilibrium competition parameters, $C_j^*$, as part of $\Delta \rho_i$, whereas previous versions of MCT shunted the $C_j^*$ to the density-dependent effects, which are then denoted by $r_i'$ (see \cite{barabas2018chesson}, Eq.19). Second, we scale resident growth rates by \textit{speed conversion factors}, whereas previous versions of MCT scaled resident growth rates by the so-called \textit{scaling factors}. Both the shunting of $C_j^*$, and the scaling factors have a very specific function: to cancel $\Delta \rho_i$. As we have argued elsewhere (\cite{johnson2022methods}), cancelling $\Delta \rho_i$ can be useful in the context of theoretical research (i.e., using simple models to derive insight), but is not recommended for "measuring coexistence" (i.e., using MCT to infer the mechanisms of coexistence in real communities). In fact, scaling factors can dramatically modulate the values of other coexistence mechanisms, potentially leading to misleading inferences about coexistence.

We scale each residents' average growth rate by $a_i / a_r$, where $a_j$ is a constant which represents the intrinsic "speed" of species $j$'s population dynamics: the capacity to quickly grow or decline.  To operationalize \textit{speed}, we typically select $a_j = 1/ \text{"generation time"}_j$, where generation time may be calculated from model parameters (\cite{CaswellHal2001Mpm:}, Section 3.5.3; \cite{bienvenu2015new}; \cite{ellner2018generation}). When the species in question do not have dramatically different generation times, it is often reasonable (and in some models, considerably simpler) to define $a_j = 1$ such that $a_i/a_r = 1$ for all $i$ and $r$. 

\subsection{Exact coexistence mechanisms}
\label{sec:Spatiotemporal coexistence mechanisms:Exact coexistence mechanisms}

The sum of small-noise coexistence mechanisms merely approximates the invasion growth rate (\eqref{sn_co_mech}). The approximation will be good if environmental fluctuations are small (see Appendix \ref{app:Deriving small-noise coexistence mechanisms:Small noise assumptions} for all assumptions), but in empirically-calibrated models there is no guarantee that the small-noise assumptions will be met. An alternative approach is to define a set of coexistence mechanisms that sum exactly to the invasion growth rate. We call these \textit{exact coexistence mechanisms} and demarcate them with the superscript "$(e)$", e.g., the exact relative nonlinearity is $\Delta N_i^{(e)}$. 

The average growth rate of species $j$ can be broken into two terms: 
\begin{equation}
    \E{t}{\log(\widetilde{\lambda}_j)} = \underbrace{\E{t}{\log(\E{x}{\lambda_j})}}_{\raisebox{.5pt}{\textcircled{\raisebox{-.9pt} {1}}}} + \underbrace{\left[\E{t}{\log(\widetilde{\lambda}_j)} - \E{t}{\log(\E{x}{\lambda_j})}\right]}_{\raisebox{.5pt}{\textcircled{\raisebox{-.9pt} {2}}}}.
\end{equation}
The first term captures the appropriate spatiotemporal average of fitness. The second term captures the effects of variation in relative density, which can be seen either by noting that  $\E{t}{\log(\widetilde{\lambda}_j)} =  \E{t}{\log(\E{x}{\lambda_j})}$ when fitness-density covariance is zero (\eqref{time_decomp_2}), or that the second term will approximate $\E{t}{\Cov{x}{\nu_j}{\lambda_j}}$ when the small-noise assumptions (Appendix \ref{app:Deriving small-noise coexistence mechanisms:Small noise assumptions}) are met. 

The first term can be further decomposed with the following schema:
\begin{equation}
    \E{t}{\log( \E{x}{\lambda})} =  \overline{\mathscr{E}_j}' +  \overline{\mathscr{C}_j}' +  \overline{\mathscr{I}_j}'
\end{equation}
\begin{equation} \label{curly_E'}
    \overline{\mathscr{E}_j}' = \E{t}{\log(\E{x}{g_j(E_j,C_j^*)})}
\end{equation}
\begin{equation} \label{curly_C'}
    \overline{\mathscr{C}_j}' = \E{t}{\log(\E{x}{g_j(E_j^*, C_j)})}
\end{equation}
\begin{equation} \label{curly_I'}
    \overline{\mathscr{I}_j}' = \E{t}{\log(\E{x}{g_j})} - \overline{\mathscr{E}_j}' - \overline{\mathscr{C}_j}'
\end{equation}
The term $\overline{\mathscr{E}_j}'$ is the main effect of the environment on the average growth rate, $ \overline{\mathscr{C}_j}' $ is the main effect of competition, and $ \overline{\mathscr{I}_j}' $ is the interaction effect between environment and competition, in analogy with a two-way ANOVA. 

The quantities $\overline{\mathscr{E}_j}'$ and $\overline{\mathscr{C}_j}'$ can be computed generically using simulation data. To compute $\overline{\mathscr{E}_j}'$, one must calculate all $\lambda_j$ while holding competition at $C_j^*$. To compute $\overline{\mathscr{C}_j}'$, one must calculate all $\lambda_j$ while holding the environment at $E_j^*$; the trick here is to still use the $C_j$ that we would have obtained had we not held the environment at $E_j^*$. To obtain these unadultered $C_j$, we first run a business-as-normal simulation whilst recording $E_j$ and $C_j$.

To calculate the exact coexistence mechanisms, our new quantities ($\overline{\mathscr{E}_j}'$,  $\overline{\mathscr{C}_j}'$, and $\overline{\mathscr{I}_j}'$)  are used in the invader--resident comparison (\eqref{inv_res_dif}) in lieu of the appropriately averaged Taylor series terms (i.e., the additive terms in \eqref{big_decomp}).

\begin{tcolorbox}[breakable, title = Formulas for exact coexistence mechanisms, subtitle style={boxrule=0.4pt,
colback=black!30!white}]

\tcbsubtitle{The invasion growth rate}

\begin{equation} \label{exact_co_mech}
    \E{t}{\log(\widetilde{\lambda}_{i})} = {\Delta E_{i}}^{(e)} + {\Delta\rho_i}^{(e)} + {\Delta\mathrm{N}_i}^{(e)} + {\Delta\mathrm{I}_i}^{(e)} + {\Delta\kappa_i}^{(e)},
\end{equation}

\tcbsubtitle{Density-independent effects}

\begin{flalign} \label{dEe}
     {\Delta E_{i}}^{(e)} & = \overline{\mathscr{E}_i}' - \frac{1}{S-1} \sum\limits_{r \neq i}^S \frac{a_i}{a_r} \overline{\mathscr{E}_r}'
\end{flalign}

\begin{flalign}
    \overline{\mathscr{E}}_{i}' & =\E{t}{\log(\E{x}{g_i(E_i,C_i^*)})}
\end{flalign}

\tcbsubtitle{Linear density-dependent effects}
      
\begin{flalign} \label{drhoe}
 {\Delta \rho_i'}^{(e)} & = \log(g_i(E_i^*, \E{x,t}{C_i})) - \frac{1}{S-1} \sum\limits_{r \neq i}^S \frac{a_i}{a_r} \log(g_i(E_r^*, \E{x,t}{C_r})) 
\end{flalign}

\tcbsubtitle{Relative nonlinearity}
      
\begin{flalign} \label{dNe}
     {\Delta N_i}^{(e)} & = \left[ \overline{\mathscr{C}_i}' - \frac{1}{S-1} \sum\limits_{r \neq i}^S \frac{a_i}{a_r} \overline{\mathscr{C}_r}' \right] - {\Delta \rho_i}^{(e)} 
\end{flalign}     

\begin{flalign}
    \overline{\mathscr{C}}_{i}' & = 
    \E{t}{\log(\E{x}{g_i(E_i^*,C_i)})}  
\end{flalign}

\tcbsubtitle{The storage effect}
      
\begin{flalign} \label{dIe}
     {\Delta I_{i}}^{(e)} & = \overline{\mathscr{I}_i}' - \frac{1}{S-1} \sum\limits_{r \neq i}^S \frac{a_i}{a_r} \overline{\mathscr{I}_r}' 
\end{flalign} 

\begin{flalign} \label{curly_I'}
    \overline{\mathscr{I}_j}' & = \E{t}{\log(\E{x}{\lambda_j})} - \overline{\mathscr{E}_j}' - \overline{\mathscr{C}_j}'  
\end{flalign}

\tcbsubtitle{Fitness-density covariance}
\begin{flalign} \label{dkappae}
    \Delta {\kappa_{i}}^{(e)} & = \left(\E{t}{\log(\widetilde{\lambda}_i)} - \E{t}{\log(\E{x}{\lambda_i})} \right) - \frac{1}{S-1} \sum\limits_{r \neq i}^S \frac{a_i}{a_r} \left(\E{t}{\log(\widetilde{\lambda}_r)} - \E{t}{\log(\E{x}{\lambda_r})} \right) \\ 
    & = \E{t}{\log(\widetilde{\lambda}_{i})} - \left( {\Delta E_{i}}^{(e)} + {\Delta\rho_i}^{(e)} + {\Delta\mathrm{N}_i}^{(e)} + {\Delta\mathrm{I}_i}^{(e)} \right)
\end{flalign} 

\end{tcolorbox}

\subsection{The space-time decomposition of small-noise coexistence mechanisms}
\label{sec:Spatiotemporal coexistence mechanisms:The space-time decomposition of small-noise coexistence mechanisms}

Ideally, we would like to take any coexistence mechanism that relies on spatiotemporal variation, and perform a \textit{space-time decomposition} to generate four additive components: the contribution of average $E_j$ and $C_j$, the contribution of spatial variation, the contribution of temporal variation to the coexistence mechanism, and the remainder (whatever is left over). For example, we would like to write the density-independent effects as $ \Delta E_{i} =  \Delta E_{i,A} + \Delta E_{i,S} + \Delta  E_{i,T} +  \Delta E_{i,R}$, with the subscripts $A$, $S$, $T$, and $R$ respectively corresponding to the average component, the space component, the time component, and the space-time interaction ($R$ stands for remainder, since the letter $I$ is already used in $\Delta I_i$ and $\overline{\mathscr{I}_i}'$).

Before decomposing entire coexistence mechanisms, we will decompose $\Var{x,t}{E_j}$, a building block of the $\Delta E_i$ coexistence mechanism. The space-time decomposition of $\Var{x,t}{E_j}$ is
\begin{equation}
\Var{x,t}{E_j} = S_j + T_j + R_j
\label{space_time_decomp}
\end{equation}
\begin{align} 
S_j = & \Var{x}{\E{t}{E_j}} \label{space term} \\ 
T_j = & \Var{t}{\E{x}{E_j}} \label{time term} \\
R_j = & \Var{x,t}{E_j} - (S_j +T_j) \label{remainder} \\
 = & \E{x}{\Var{t}{E_j}} -  \Var{t}{\E{x}{E_j}} \nonumber \\
= & \E{t}{\Var{x}{E_j}} -  \Var{x}{\E{t}{E_j}}. \nonumber
\end{align}%
The last two expressions for $R_j$ are obtained using the law of total variance. A close examination confirms our space-time decomposition satisfies some minimal requirements: $S_j = 0$ when there is no spatial variation in $E_j$, $T_j = 0$ when there is no temporal variation, and $R_j = 0$ when there is either no spatial or temporal variation.

The components of the space-time decomposition of $\Var{x,t}{E_j}$ can be thought of as differences between hypothetical worlds in which spatial and/or temporal variation has been turned on or off. For example, the space term, $S_j$, is the difference between the variance of $E_j$ in a world where temporal variation has been turned off (by setting $E_j(x,t)$ to $\E{t}{E_j}$) leaving only spatial variation, and the variance of $E_j$ in a reference world where both spatial and temporal variation have turned off (which is necessarily zero). Adding only spatial variation to the reference state of "no variation" gives the main effect of spatial variation. The interaction effect of spatial and temporal variation is the marginal effect of turning on both spatial and temporal variation; it is the extent to which the combination of spatial and temporal variation exceeds the sum of its parts, which is why the interaction term (\eqref{remainder}) involves subtracting both main effects.

Our talk of "hypothetical worlds" and "turning off variation" may give our space-time decomposition a speciously ad hoc aura. In Appendix \ref{app:Justification of the space-time decomposition}, we justify our space-time decomposition by 1) showing that the results it gives in a toy model accords with intuition, and 2) using the philosophical literature to show that our decomposition results in terms that can be interpreted as the \textit{causal effects} of spatial and temporal variation.

In \eqref{space_time_decomp}--\ref{remainder}, we defined the space-time decomposition of $\Var{x,t}{E_j}$. The other variance/covariance terms featured in the small-noise coexistence mechanisms (i.e., $\Var{x,t}{C_j}$ and $\Cov{x,t}{E_j}{C_j}$) can be decomposed in analogous fashion, by turning on/off $E_j$ and $C_j$ in tandem. To obtain the space-time decomposition of the small-noise coexistence mechanisms, we propagate the small-noise decompositions of $\Var{x,t}{E_j}$, $\Var{x,t}{C_j}$, and $\Cov{x,t}{E_j}{C_j}$ though the expressions for the small-noise coexistence mechanisms. (\eqref{dE}--\ref{dkappa}). For example, since variance in $E_j$ is the purview of the density-independent effects ($\Delta E_{i}$), and because the space-component of $\Var{x,t}{E_j}$ is $\Var{x}{\E{t}{E}}$, it follows that all terms involving $\Var{x}{\E{t}{E}}$ will belong to $\Delta E_{i,S}$, the space-component of the density-independent effects. 

All averages over space and time are shunted into the "Average" components of the space-time time decomposition, denoted with the subscript $A$. Note that relative nonlinearity ($\Delta N_i$) has no average component because the average effect of $C_j$ is captured in the linear density-dependent effects ($\Delta \rho_i$). Also note that the average component of the storage effect ($\Delta I_{i,A}$) equals zero, since the covariance between two constants is always zero. Fitness-density covariance could technically be decomposed as $\Delta \kappa = \Delta \kappa_{i,S} + \Delta \kappa_{i,R}$, but we choose not to decompose $\Delta \kappa$ on the grounds that it is  inherently a spatial coexistence mechanism. 

\begin{tcolorbox}[breakable, title = Formulas for space-time decomposition of small-noise coexistence mechanisms, subtitle style={boxrule=0.4pt,
colback=black!30!white}]

\tcbsubtitle{Density-independent effects}

\begin{equation} 
 \Delta E_{i} =  \Delta E_{i,A} +  \Delta E_{i,S} +  \Delta E_{i,T} +  \Delta E_{i,R}
\end{equation}

\begin{flalign}
\label{dr EA}
     \Delta E_{i,A} & = \alpha_{i}^{(1)} \E{x,t}{E_i - E_{i}^{*}}  - \frac{1}{S-1} \sum\limits_{r \neq i}^S \frac{a_i}{a_r} \alpha_r^{(1)} \E{x,t}{E_r - E_{r}^{*}} & \\
\label{dDelta ES}
     \Delta E_{i,S} & = \frac{1}{2} \alpha_i^{(2)} \Var{x}{\E{t}{E_i}} - \frac{1}{S-1} \sum\limits_{r \neq i}^S \frac{a_i}{a_r}  \frac{1}{2} \alpha_r^{(2)}\Var{x}{\E{t}{E_r}}  \\
\label{dr'T}
     \Delta E_{i,T} & = \frac{1}{2} \left( \alpha_i^{(2)} - \alpha_i^{(1)^{2}} \right) \Var{t}{\E{x}{E_i}}
   - \frac{1}{S-1} \sum\limits_{r \neq i}^S \frac{a_i}{a_r}  \frac{1}{2} \left( \alpha_r^{(2)} - \alpha_i^{(1)^{2}} \right) \Var{t}{\E{x}{E_r}} & \\
\label{dr'R}
     \Delta E_{i,R} & = \frac{1}{2} \alpha_i^{(2)} \left[ \E{t}{\Var{x}{E_i}} - \Var{x}{\E{t}{E_i}} \right]  & \\
    & - \frac{1}{S-1} \sum\limits_{r \neq i}^S \frac{a_i}{a_r}  \frac{1}{2} \alpha_r^{(2)} \left[ \E{t}{\Var{x}{E_r}} - \Var{x}{\E{t}{E_r}} \right] \nonumber & \\
    & = \frac{1}{2} \alpha_i^{(2)} \left[ \E{x}{\Var{t}{E_i}} - \Var{t}{\E{x}{E_i}} \right] \nonumber & \\
    & - \frac{1}{S-1} \sum\limits_{r \neq i}^S \frac{a_i}{a_r}  \frac{1}{2} \alpha_r^{(2)} \left[ \E{x}{\Var{t}{E_r}} - \Var{t}{\E{x}{E_r}} \right] \nonumber &
\end{flalign}

\tcbsubtitle{Linear density-dependent effects}
      
\begin{flalign}
   \Delta\rho_i = \beta_i^{(1)} \E{x,t}{C_i - C_{i}^{*}}  - \frac{1}{S-1}  \sum\limits_{r \neq i}^S \frac{a_i}{a_r} \beta_r^{(1)} \E{x,t}{C_r - C_{r}^{*}}
\end{flalign}      

\tcbsubtitle{Relative nonlinearity}
      
\begin{equation} 
\Delta N_{i} = \Delta N_{i,S} + \Delta N_{i,T} + \Delta N_{i,R},
\end{equation}

\begin{flalign}
\label{dNS}
    \Delta N_{i,S} & = \frac{1}{2} \beta_i^{(2)}\Var{x}{\E{t}{C_i}} - \frac{1}{S-1}  \sum\limits_{r \neq i}^S \frac{a_i}{a_r}  \frac{1}{2} \beta_r^{(2)}\Var{x}{\E{t}{C_r}}  & \\
\label{dNT}
    \Delta N_{i,T} & = \frac{1}{2} \left( \beta_i^{(2)} - \beta_i^{(1)^{2}} \right) \Var{t}{\E{x}{C_i}}
   - \frac{1}{S-1} \sum\limits_{r \neq i}^S \frac{a_i}{a_r}  \frac{1}{2} \left( \beta_r^{(2)} - \beta_i^{(1)^{2}} \right) \Var{t}{\E{x}{C_r}} & \\
\label{dIR}
    \Delta N_{i,R} & = \frac{1}{2} \beta_i^{(2)} \left[ \E{t}{\Var{x}{C_i}} - \Var{x}{\E{t}{C_i}} \right]  & \\ 
    & -\frac{1}{S-1}  \sum\limits_{r \neq i}^S \frac{a_i}{a_r}  \frac{1}{2} \beta_r^{(2)} \left[ \E{t}{\Var{x}{C_r}} - \Var{x}{\E{t}{C_r}} \right] \nonumber & \\
    & = \frac{1}{2} \beta_i^{(2)} \left[ \E{x}{\Var{t}{C_i}} - \Var{t}{\E{x}{C_i}} \right] \nonumber & \\ 
    & - \frac{1}{S-1} \sum\limits_{r \neq i}^S \frac{a_i}{a_r}  \frac{1}{2} \beta_r^{(2)} \left[ \E{x}{\Var{t}{C_r}} - \Var{t}{\E{x}{C_r}} \right] \nonumber & 
\end{flalign}

\tcbsubtitle{The storage effect}
      
\begin{equation}
\Delta I_{i} = \Delta I_{i,A} + \Delta I_{i,S} + \Delta I_{i,T} + \Delta I_{i,R}
\end{equation}
 
\begin{flalign}
\label{dIA}
    \Delta I_{i,A} & = 0 \\
\label{dIS}
    \Delta I_{i,S} & = \zeta_i \Cov{x}{\E{t}{E_i}}{\E{t}{C_i}} - \frac{1}{S-1} \sum\limits_{r \neq i}^S \frac{a_i}{a_r}  \zeta_r \Cov{x}{\E{t}{E_r}}{\E{t}{C_r}} & \\
\label{dIT}
    \Delta I_{i,T} & = \left(\zeta_i - \alpha_i^{(1)}\beta_i^{(1)}\right) \Cov{t}{\E{x}{E_i}}{\E{x}{C_i}} \nonumber & \\
   & - \frac{1}{S-1} \sum\limits_{r \neq i}^S \frac{a_i}{a_r}  \left(\zeta_r - \alpha_r^{(1)}\beta_r^{(1)}\right) \Cov{t}{\E{x}{E_r}}{\E{x}{C_r}} & \\
\label{dIR}
    \Delta I_{i,R} 
    & =  \left[ \zeta_i\left( \E{t}{\Cov{x}{E_i}{C_i}} - \Cov{x}{\E{t}{E_i}}{\E{t}{C_i}}\right) \right]  & \\ 
    & - \frac{1}{S-1} \sum\limits_{r \neq i}^S \frac{a_i}{a_r}  \left[ \zeta_r\left( \E{t}{\Cov{x}{E_r}{C_r}} - \Cov{x}{\E{t}{E_r}}{\E{t}{C_r}}\right) \right] \nonumber & \\
    & = \left[ \zeta_i\left( \E{x}{\Cov{t}{E_i}{C_i}} - \Cov{t}{\E{x}{E_i}}{\E{x}{C_i}}\right) \right] \nonumber & \\ 
    & - \frac{1}{S-1} \sum\limits_{r \neq i}^S \frac{a_i}{a_r}  \left[ \zeta_r\left( \E{x}{\Cov{t}{E_r}{C_r}} - \Cov{t}{\E{x}{E_r}}{\E{x}{C_r}}\right) \right] \nonumber &
\end{flalign}

\end{tcolorbox}

\subsection{The space-time decomposition of exact coexistence mechanisms}
\label{sec:Spatiotemporal coexistence mechanisms:The space-time decomposition of exact coexistence mechanisms}

In this section, we will describe how the space-time decomposition of the exact coexistence mechanisms can be computed using data from simulations. Our exposition is focused on the storage effect because it is the most difficult exact coexistence mechanism to quantify, but we will give formulae for the other coexistence mechanisms at the end of this section.

\textcite{ellner2016quantify} showed how simulations could be used to calculate the exact temporal storage in a model with only temporal variation. Their procedure can be naturally extended to models with spatiotemporal variation:

\begin{enumerate}
    \item Simulate the model. For each species, record a matrix of $E_j(x,t)$'s and a matrix of $C_j(x,t)$'s with each row corresponding to a location in space, and each column corresponding to a point in time. Call these matrices $\boldsymbol{E_j}$ and $\boldsymbol{C_j}$.
    
    \item For each species, shuffle the elements of $\boldsymbol{E_j}$. That is, fill in a matrix with equivalent dimensions by randomly sampling without replacement from the flattened $\boldsymbol{E_j}$. Call this new matrix $\boldsymbol{E_j}^{\#}$. Shuffling (i.e., randomly sampling without replacement) destroys the covariance between environment and competition (as well as any higher order mixed moments) that is integral to the storage effect.
    
    \item For each species, estimate the $EC$ interaction effect as 
$ \overline{\mathscr{I}_j}' = \E{t}{\log(\E{x}{g_j(\boldsymbol{E_j}, \boldsymbol{C_j})})} - \E{t}{\log(\E{x}{g_j(\boldsymbol{E_j}^\#, \boldsymbol{C_j})})}$. Note here that we are averaging finite rates of increase across patches instead of individuals. This ensures that our estimate of ${\Delta I_{i}}^{(e)}$ does not include any bit of growth rate that can be contributed to the fitness-density covariance, $\Delta \kappa_i$. 

\item Calculate the exact storage effect as ${\Delta I_{i}}^{(e)} = \overline{\mathscr{I}_i}' - \sum_{r \neq i}^S \frac{a_i}{a_r}  \overline{\mathscr{I}_j}'$. 

\end{enumerate}

\posscite{ellner2016quantify} critical idea --- shuffling an archive of environmental states --- can also be utilized to calculate the space-time decomposition of the exact storage effect. To illustrate, we will discuss how one may calculate the space component of the precursor to the exact storage effect: $\mathscr{I}_{j,S}'$. To measure the the causal effect of spatial covariation, we must compare a hypothetical world with only spatial variation to a (reference) hypothetical world with only spatial variation and no $EC$ covariation. We obtain the hypothetical world with only spatial variation by squashing temporal variation, i.e., by setting $E_j(x,t)$ to $\E{t}{E_j(x)}$ and setting $C_j(x,t)$ to $\E{t}{C_j(x)}$. This produces the growth rate $\log(\E{x}{g_j(\E{t}{E_j},\E{t}{C_j})})$. We obtain the hypothetical world with no temporal variation and no spatial covariation by squashing temporal variation just as we did before, and then shuffling the vector of $\E{t}{E_j(x)}$. This produces the growth rate $\log(\E{x}{g_j(\E{t}{E_j}^\#,\E{t}{C_j})})$. The effects of spatial covariation (and higher order mixed moments) on species $j$'s growth rate is simply the difference between the growth rates corresponding to the two hypothetical worlds. Put into symbols, we say that $\overline{\mathscr{I}}_{j,S}'  = \log(\E{x}{g_j(\E{t}{E_j},\E{t}{C_j})})  - \log(\E{x}{g_j(\E{t}{E_j}^\#,\E{t}{C_j})})$.

Instead of writing out steps for quantifying every space-time component of every exact coexistence mechanism, we will provide formulas that indicate how simulated data are to be used. Of notable importance to the storage effect is the previously introduced shuffle operator, denoted by the superscript $\#$, which indicates that the elements of a matrix or vector are to be shuffled, i.e., randomly sampled without replacement. Note that the precursor to the "average" component of the storage effect, ${\Delta I_{i,A}}^{(e)}$, is not necessarily zero (as it was in the analogous small-noise expression) though it should be small in the limit of small-noise. 

\begin{tcolorbox}[breakable, title = Formulas for space-time decomposition of exact coexistence mechanisms, subtitle style={boxrule=0.4pt,
colback=black!30!white}]

\tcbsubtitle{Density-independent effects}

\begin{equation} \label{ri'e_st_decomp}
 {\Delta E_{i}}^{(e)} =  {\Delta E_{i,A}}^{(e)} +  {\Delta E_{i,S}}^{(e)} +  {\Delta E_{i,T}}^{(e)} +  {\Delta E_{i,R}}^{(e)}
\end{equation}

\begin{flalign}
     {\Delta E_{i,A}}^{(e)} & = \overline{\mathscr{E}}_{i,A}' - \frac{1}{S-1} \sum\limits_{r \neq i}^S \frac{a_i}{a_r} \overline{\mathscr{E}}_{r,A}' & \\
     {\Delta E_{i,S}}^{(e)} & = \overline{\mathscr{E}}_{i,S}' - \frac{1}{S-1} \sum\limits_{r \neq i}^S \frac{a_i}{a_r} \overline{\mathscr{E}}_{r,S}' & \\
     {\Delta E_{i,T}}^{(e)} & = \overline{\mathscr{E}}_{i,T}' - \frac{1}{S-1} \sum\limits_{r \neq i}^S \frac{a_i}{a_r} \overline{\mathscr{E}}_{r,T}' & \\   
     {\Delta E_{i,R}}^{(e)} & = \overline{\mathscr{E}}_{i,R}' - \frac{1}{S-1} \sum\limits_{r \neq i}^S \frac{a_i}{a_r} \overline{\mathscr{E}}_{r,R}' &
\end{flalign}

\begin{flalign}
\label{curly_E}
    \overline{\mathscr{E}}_{j}' & = 
    \E{t}{\log(\E{x}{g_j(E_j,C_j^*)})} & \\
\label{curly_E_A}
    \overline{\mathscr{E}}_{j,A}' & = 
    \log(g_j(\E{x,t}{E_j},C_j^*)) & \\
\label{curly_E_S}
    \overline{\mathscr{E}}_{j,S}' & = \log(\E{x}{g_j(\E{t}{E_j},C_j^*)}) - \overline{\mathscr{E}}_{j,A}' & \\
\label{curly_E_T}
    \overline{\mathscr{E}}_{j,T}' & = \E{t}{\log(g_j(\E{x}{E_j},C_j^*))} - \overline{\mathscr{E}}_{j,A}' & \\    
\label{curly_E_R}
    \overline{\mathscr{E}}_{j,R}' & =  \overline{\mathscr{E}}_{j}' - \left( \overline{\mathscr{E}}_{j,A}' + \overline{\mathscr{E}}_{j,S}' + \overline{\mathscr{E}}_{j,T}' \right)      
\end{flalign}

\tcbsubtitle{Linear density-dependent effects}

\begin{equation} \label{drho^e_st_decomp}
 {\Delta \rho_{i}}^{(e)} = \overline{\mathscr{C}}_{i,A}' - \frac{1}{S-1} \sum\limits_{r \neq i}^S \frac{a_i}{a_r} \overline{\mathscr{C}}_{r,A}'
\end{equation}

\begin{flalign}
\label{curly_C_A1}
    \overline{\mathscr{C}}_{j,A}' & = 
    \log(g_i(E_i^*, \E{x,t}{C_i})) &
\end{flalign}

\tcbsubtitle{Relative nonlinearity}

\begin{equation} \label{dN^e_st_decomp}
{\Delta N_{i}}^{(e)} = {\Delta N_{i,S}}^{(e)} + {\Delta N_{i,T}}^{(e)} + {\Delta N_{i,R}}^{(e)},
\end{equation}

\begin{flalign}
\label{dN_S^e}
     {\Delta N_{i,S}}^{(e)} & = \overline{\mathscr{C}}_{i,S}' - \frac{1}{S-1} \sum\limits_{r \neq i}^S \frac{a_i}{a_r} \overline{\mathscr{C}}_{r,S}' & \\
\label{dN_T^e}
     {\Delta N_{i,T}}^{(e)} & = \overline{\mathscr{C}}_{i,T}' - \frac{1}{S-1} \sum\limits_{r \neq i}^S \frac{a_i}{a_r} \overline{\mathscr{C}}_{r,T}' & \\   
\label{dN_R^e}
     {\Delta N_{i,R}}^{(e)} & = \overline{\mathscr{C}}_{i,R}' - \frac{1}{S-1} \sum\limits_{r \neq i}^S \frac{a_i}{a_r} \overline{\mathscr{C}}_{r,R}'  & 
\end{flalign}

\begin{flalign}
\label{curly_C}
    \overline{\mathscr{C}}_{j}' & = 
    \E{t}{\log(\E{x}{g_j(E_j^*,C_j)})} & \\
\label{curly_C_A}
    \overline{\mathscr{C}}_{j,A}' & = 
    \log(g_j(E_j^*, \E{x,t}{C_j})) & \\
\label{curly_C_S}
    \overline{\mathscr{C}}_{j,S}' & = \log(\E{x}{g_j(E_j^*,\E{t}{C_j})}) - \overline{\mathscr{C}}_{j,A}'  & \\
\label{curly_C_T}
    \overline{\mathscr{C}}_{j,T}' & = \E{t}{\log(g_j(E_j^*,\E{x}{C_j}))} - \overline{\mathscr{C}}_{j,A}'  & \\    
\label{curly_C_R}
    \overline{\mathscr{C}}_{j,R}' & =  \overline{\mathscr{C}}_{j}' - \left( \overline{\mathscr{C}}_{j,A}' + \overline{\mathscr{C}}_{j,S}' + \overline{\mathscr{C}}_{j,T}' \right)   &    
\end{flalign}

\tcbsubtitle{The storage effect}

\begin{equation} \label{dI^e_st_decomp}
{\Delta I_{i}}^{(e)} = {\Delta I_{i,A}}^{(e)} + {\Delta I_{i,S}}^{(e)} + {\Delta I_{i,T}}^{(e)} + {\Delta I_{i,R}}^{(e)},
\end{equation}

\begin{flalign}
\label{dI_A^e}
     {\Delta I_{i,A}}^{(e)} & = \overline{\mathscr{I}}_{i,A}' - \frac{1}{S-1} \sum\limits_{r \neq i}^S \frac{a_i}{a_r} \overline{\mathscr{I}}_{r,A}' & \\
\label{dI_S^e}
     {\Delta I_{i,S}}^{(e)} & = \overline{\mathscr{I}}_{i,S}' - \frac{1}{S-1} \sum\limits_{r \neq i}^S \frac{a_i}{a_r} \overline{\mathscr{I}}_{r,S}' & \\
\label{dI_T^e}
     {\Delta I_{i,T}}^{(e)} & = \overline{\mathscr{I}}_{i,T}' - \frac{1}{S-1} \sum\limits_{r \neq i}^S \frac{a_i}{a_r} \overline{\mathscr{I}}_{r,T}' & \\   
\label{dI_R^e}
     {\Delta I_{i,R}}^{(e)} & = \overline{\mathscr{I}}_{i,R}' - \frac{1}{S-1} \sum\limits_{r \neq i}^S \frac{a_i}{a_r} \overline{\mathscr{I}}_{r,R}' &   
\end{flalign}

\begin{flalign}
\label{curly_I}
    \overline{\mathscr{I}}_{j}' & = \E{t}{\log(\E{x}{g_j(E_j,C_j)})} - \left( \overline{\mathscr{E}}_{j}' + \overline{\mathscr{C}}_{j}' \right) \\
    & \approx \E{t}{\log(\E{x}{g_j(E_j,C_j)})} - \E{t}{\log(\E{x}{g_j(E_j^{\#},C_j)})} & \\
    \label{curly_j_A}
    \overline{\mathscr{I}}_{j,A}' & = \log(g_j( \E{x,t}{E_j},\E{x,t}{C_j})) - \left( \overline{\mathscr{E}}_{j,A}' + \overline{\mathscr{C}}_{j,A}' \right) & \\
   \label{curly_j_S}
    \overline{\mathscr{I}}_{j,S}' & = \log(\E{x}{g_j(\E{t}{E_j},\E{t}{C_j})})  - \log(\E{x}{g_j(\E{t}{E_j}^\#,\E{t}{C_j})})  & \\
\label{curly_j_T}
    \overline{\mathscr{I}}_{j,T}' & = \E{t}{\log(g_j(\E{x}{E_j},\E{x}{C_j)}))} - \E{t}{\log(g_j(\E{x}{E_j}^\#,\E{x}{C_j}))} & \\  
\label{curly_j_R}
    \overline{\mathscr{I}}_{j,R}' & =  \overline{\mathscr{I}}_{j}' - \left( \overline{\mathscr{I}}_{j,A}' + \overline{\mathscr{I}}_{j,S}' + \overline{\mathscr{I}}_{j,T}' \right) &      
\end{flalign}

\end{tcolorbox}

\section{Computational tricks for measuring invasion growth rates in particular classes of models}
\label{sec:Extended discussion of invasion growth rates, including computational tricks}

We have given formulas for computing coexistence mechanisms, but the components of the those formulas (e.g., $E_j$ and $C_j$) must be measured in a specific context. Specifically, the invasion growth rate and coexistence mechanisms must be measured in the context where 1) the invader's environment (which includes the resident species) has attained its limiting dynamics, and 2) the invader has attained its quasi-steady spatial distribution.

Here, "the invader's environment" does not refer to the environmental parameter $E_i$, but rather all variables that influence the invader's per capita growth rate (e.g., resident densities, resources, temperature). Previous expositions of MCT required that the invader's environment be an ergodic stationary stochastic process (\cite[p.~236]{Chesson1994}). This assumption is convenient because ergodicity implies that initial conditions are irrelevant, and stationarity allows the long-term average (inherent in the invasion growth rate) to be replaced with the expectation over the stationary distribution of the state of the invader's environment; as we will see, there are several well-established tricks for calculating stationary distributions. However, requiring a stationary distribution excludes any models where parameters change over time, including models with seasonality and models that track weather patterns. Instead, we only a require that the invader's environment has a unique, asymptotic, time-average distribution (\cite{glynn1998independent}), which only excludes models with unidirectional environmental trends.

In many ecological models, the time-average distribution of invader's environment is a \textit{stationary distribution} (\cite{Nisbet1982Modelling}). In homogeneous (i.e., time-invariant) Markov chain models with a finite number of states, the stationary distribution can be computed as the dominant eigenvector of the \textit{transition probability matrix} or the \textit{generator matrix} (terminology changes depending on whether the model is in discrete time or continuous time; \cite[p.~67]{allen2010introduction}). When the state space is the natural numbers (i.e., there are a countable but infinite number of states), one may approximate the stationary distribution as the dominant eigenvector of a truncated transition probability matrix (or generator matrix) where rows and columns corresponding to states of high abundance have been removed (\cite[p.~128-129]{allen2010introduction}). Alternatively, one may obtain an approximate stationary distribution using the Wentzel–Kramers–Brillouin (WKB) approximation (\cite{assaf2010extinction}; \cite{pande2020taming}). For models that take the form of stochastic differential equations, the stationary distribution can be obtained by solving a second order differential equation (\cite{karlin1981second}, ch. 15.3). Alternatively, one may obtain an approximate stationary distribution by finding the minimum action of a path integral (\cite{Chow2015}; \cite{kamenev2008colored}). However, because the state space of all species' densities/abundances increases exponentially with the number of species, the computation time for all the aforementioned methods scales exponentially with the number of species under consideration. 

For models with many species, or models where the notion of \textit{stationarity} is not appropriate, one may have to take a brute-force approach: simulate a model forward in time, recording the frequency distribution of different states after a sufficiently long burn-in period. To determine the length of the burn-in period, one may simply "eye-ball" a time series plot, perhaps selecting $2\times$ the time it takes for the residents to attain typical densities. When one must obtain the time-average distribution for many different parameter combinations, the "eye-ball" approach becomes impractical. Instead, one can employ heuristic tests for determining the length of the burn-in period (see \cite{caswell1993ecological}; \cite{hiebeler2011pair}).

MCT assumes that all populations have infinite population sizes; otherwise, the invader could go extinct before it experiences a representative collection of environmental states, in which case the invasion growth rate would depend on the initial conditions of the invader's environment. Because the resident species can also go extinct in finite-population models, the concept of the stationary distribution can be replaced with the quasi-stationary distribution (QSD): the distribution of resident densities conditioned on non-extinction. In single-resident birth-death models, there is iterative numerical procedure for finding the quasi-stationary distribution (\cite[p.~183--184]{Nisbet1982Modelling}). Unlike the stationary distribution, the QSD cannot be computed with naive simulation. The problem is that a simulation must run for a long time in order for the frequency distribution to converge, but the longer the simulation, the more likely extinction is. One solution is the Fleming-Voit method (\cite{ferrari2007quasi}; \cite{blanchet2016analysis}), where a number simulations are run in parallel so that extinct simulations can be restarted with initial conditions equal to the state of one of the other simulations. A similar method restarts extinct simulations by drawing randomly from an archive of past states (\cite{groisman2012simulation}). 

To avoid simulations, one may \textit{approximate} the QSD by analyzing an auxiliary model. This auxiliary model is exactly like the original model, except either 1) each transition from a non-zero state to the zero state (i.e., extinction), has probability equal to zero (\cite[p.~27]{PielouE.C.1969Aitm}; \cite[p.~127]{allen2010introduction}), or 2) one individual is immortal for all time (\cite{weiss1971asymptotic}; \cite{norden1982distribution}). The stationary distribution of the auxiliary model (computed using the methods in the previous paragraphs) is an approximation of the quasi-stationary distribution of the original model. The auxiliary model \#1 leads to better results for populations with long mean extinction times, whereas the auxiliary model \#2 leads to better results for populations with short mean extinction times (\cite{naasell2001extinction}; \cite{kryscio1989extinction}).

A unique challenge in spatiotemporal models (with either infinite or finite populations) is determining the \textit{quasi-steady spatial distribution} of the invader, not to be confused with the previously discussed quasi-stationary distribution of resident densities. To accurately measure the invasion growth rate, one must inoculate the invader and then wait until it has attained its natural spatial distribution. However, the longer one waits for the the invader to attain this distribution, the larger the invader population becomes (assuming a positive invasion growth rate and barring stochastic extinction), leading to inaccurate measurements of the invasion growth rate. One hopes that the dynamics of spatial correlations operate on a much faster timescale than the dynamics of total density, such that a quasi-steady spatial distribution (i.e., a second-order stationary and isotropic process; \cite{cressie2015statistics}) is attained long before the total density changes too much. The requisite time-scale separation can be verified by plotting spatial correlations against total density (as in \cite{le2003adaptive}, Fig. 7). Analytical expressions for the quasi-steady distribution are only available in simple spatially implicit models (see Appendix \ref{app:Deriving the small-noise fitness-density covariance for the spatiotemporal lottery model} for a worked example) or in simple spatially explicit models with the help of \textit{pair approximation} (\cite{ferriere2001invasion}). 

In more complex models, simulation experiments are needed to compute the quasi-steady spatial distribution of the invader. After virtually inoculating the invader species and waiting through a sufficiently long burn-in period, one can begin measuring the invasion growth rate. If the regional invader population density exceeds a user-specified ceiling (i.e., the invader becomes common), then the simulation can be restarted. Indeed, this general strategy can be used to compute other kinds of quasi-steady distributions, such as the invader's stable-age distribution. In finite population models, the invader may go extinct. To circumvent this problem, one may apply the previously discussed Fleming-Voit method (\cite{ferrari2007quasi}; \cite{blanchet2016analysis}). 

\section{Example: the spatiotemporal lottery model}
\label{sec:Example: the spatiotemporal lottery model}

\begin{figure}[h]
      \includegraphics[scale = 0.5]{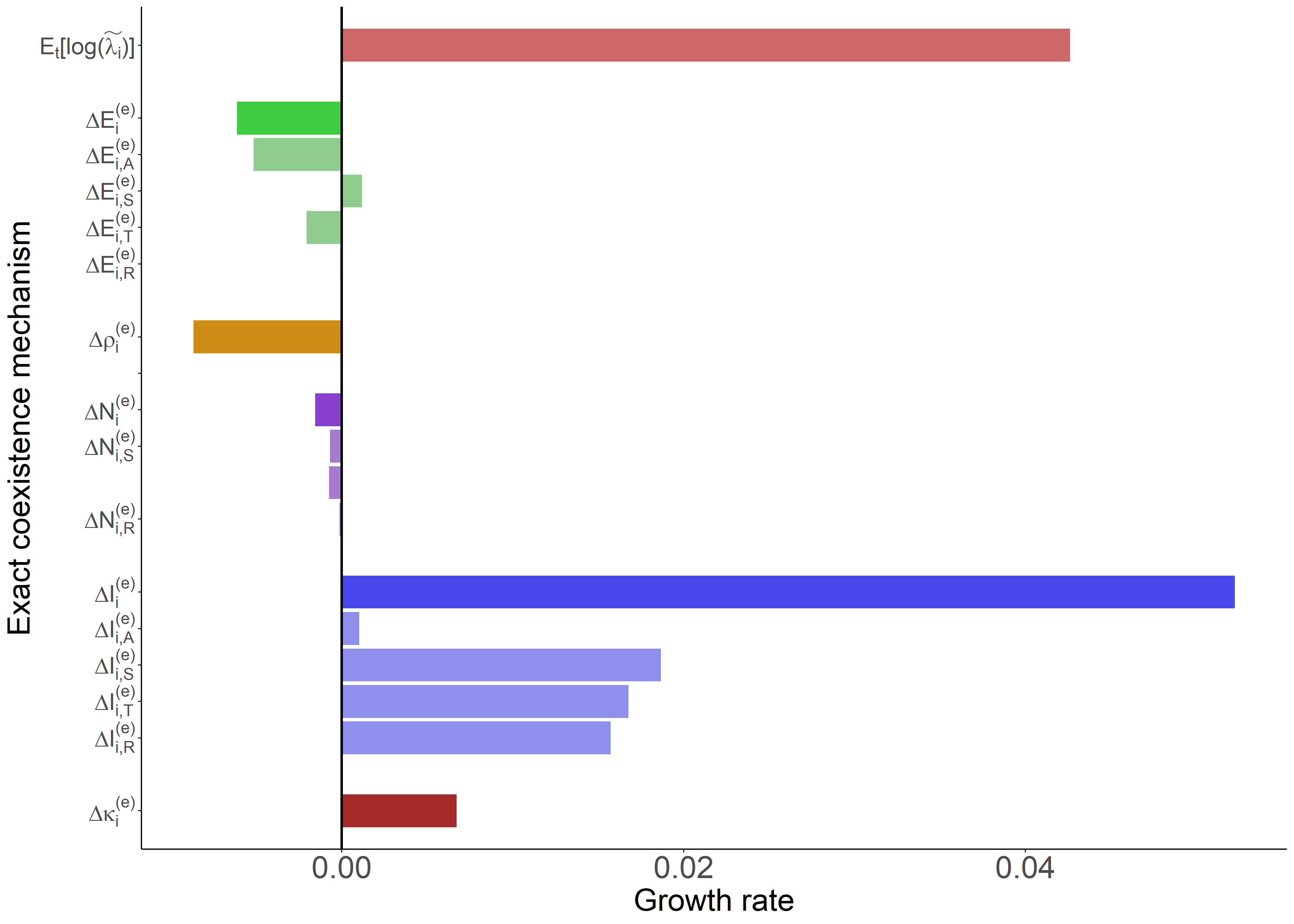}
  \caption{Exact coexistence mechanisms in the spatiotemporal lottery model with 3 species. Coexistence can be attributed to the storage effect and fitness-density covariance. Parameter values and simulation code can be found in {\fontfamily{qcr}\selectfont lottery\_model\_example.R}.}
  \label{fig:lottery coef plot}
\end{figure}

To give readers a sense of how Spatiotemporal MCT may be used in practice, we analyze the lottery model (\cite{chesson1981environmentalST}; \cite{Chesson1994}) with spatiotemporal fluctuations. The lottery model is one of the simplest models that features fluctuation-dependent coexistence mechanisms, and has thus become a canonical model in theoretical ecology. We derive analytical expressions for the special case of two species with similar demographic parameters (\eqref{dE_LM}--\eqref{dkappa_LM}). Additionally, we compute exact coexistence mechanisms in a three-species system with dissimilar parameters (Fig. \ref{fig:lottery coef plot}).

Imagine several fish species inhabiting territories on a coral reef. During each time-step, an individual of species $j$ produces $\xi_j(x,t)$ larvae; per capita larval production fluctuates over space and time. The remaining life history is very simple. Adult fish die with the density-independent probability $\delta_j$. Within a single patch, the larvae inherit the empty territories with a per-larva recruitment probability equal to the number of empty sites, divided by the total number of larvae. The remaining larvae perish. Note that "empty territories" and "total larvae" here are patch-specific quantities; so far, we have only described local population dynamics. The uniform per-larva probability of recruitment is the reason that this model is called the lottery model (\cite{Sale1977}). 

If there are $S$ species, the local dynamics of the lottery model can be encoded in a $S$-dimensional difference equation:

\begin{equation} \label{lottery}
   \lambda_j(x,t) =  \overbrace{1-\delta_j}^{\text{survival prob.}} +  \; \overbrace{\xi_{j}(x,t)}^{{\scriptstyle \text{per capita fecundity}}}   \left[\rule{0cm}{1.25cm}\right. \frac{ \overbrace{ \sum \limits_{j \neq i}^{S} \delta_j n_{j}(x,t)}^{\text{open territories}}}{ \underbrace{\sum \limits_{j \neq i}^{S} \xi_{j}(x,t) n_{j}(x,t)}_{ {\scriptstyle \text{total larvae}}}} \left.\rule{0cm}{1.25cm}\right] , 
\end{equation}

Selecting $E_j = \log(\xi_j)$ and $C = \log( \frac{\sum \limits_{j \neq i}^{S} \xi_{j} n_{j} }{\sum \limits_{j \neq i}^{S} \delta_j n_{j}})$, the local finite rate of increase takes the simple form,

\begin{equation} 
   g_{j}(E_j(x,t), C_j(x,t)) = 1 - \delta_j + \exp{E_j(x,t) - C(x,t)}.
\end{equation}

Both species share the same equilibrium competition parameter, $C^* = \frac{1}{S} \sum_{i=1}^{S} \E{x,t}{C_{i}}$, which is the average competition experienced by the invader, averaged over all species acting as the invader. This equilibrium competition parameter fixes the species-specific equilibrium environmental parameter at  $E_j^* = \log(\delta_j) + C^*$. With the equilibrium parameters in hand, we can now compute the Taylor series coefficients for the small-noise coexistence mechanisms: we find that $\alpha_j^{(1)} = \delta_j$, $\beta_j^{(1)} = -\delta_j$, $\alpha_j^{(2)} = \delta_j$, $\beta_j^{(2)} = \delta_j$, $\zeta_j = -\delta_j$. 

In the second segment of each time-step, after local growth occurs, a fraction of individuals, $q_j$, are retained at site $x$ while the $p_j = 1-q_j$ fraction of dispersing individuals are distributed evenly across all $K$ patches. This particular form of dispersal dynamics, which we may call \textit{local retention with global dispersal}, is easy to simulate and is analytically tractable. The full dynamics of species $j$ can now be written as 

\begin{equation} 
  N_j(x,t+1) =  q_j N_j(x,t) g_j(E_j(x,t), C_j(x,t))+ \frac{1 - q_j}{K} \sum \limits_{s=1}^{K} N_j(s,t) g_j(E_j(s,t), C_j(s,t)).
\end{equation}

Finally, we must describe the structure of environmental variation. The environmental parameter, $E_j(x,t)$, is the sum of a patch effect $a(x)$, a time effect $b(t)$, and their interaction, which is scaled by the interaction coefficient $\theta_j$:
\begin{equation} \label{lottery_noise}
    E_j(x,t) = a_j(x) + b_j(t) + \theta_j a_j(x) b_j(t)
\end{equation}
For simplicity, $a_j(x)$ and $b_j(t)$ are independently drawn from normal distributions with standard deviations $\sigma_j^{(x)}$ and  $\sigma_j^{(t)}$, respectively. There are no spatial or temporal autocorrelations, but there are cross-species correlations. The correlation between $a_j(x)$ and $a_k(x)$ is $\phi_{jk}^{(x)}$, and the correlation between $b_j(t)$ and $b_k(t)$ is $\phi_{jk}^{(t)}$. Under the small-noise assumptions of MCT, the term $\theta_j a_j(x) b_j(t)$ will become negligibly small when squared, and thus the remainder component of the space-time decomposition will be zero. For purely illustrative purposes, we will assume that $\theta_j = \mathcal{O}(\sigma^{-1})$, as this allows us to obtain a non-zero remainder component while still keeping the simple form of \eqref{lottery_noise}.

To ensure that species with fast life-cycles do not dominate the invader--resident comparison, we multiply the residents' growth rates by the speed conversion factors (Section \ref{sec:Spatiotemporal coexistence mechanisms:Small-noise coexistence mechanisms}; \cite{johnson2022methods}). The standard way to operationalize population-dynamical speed is as the reciprocal of generation time, which is equal to $\delta_j$ in the lottery model (the waiting time till death follows a geometric distribution with mean $1/\delta_j$). This makes the speed conversion factors $a_i / a_r = \delta_i / \delta_r$.

We now analyze a particularly simple case of the spatiotemporal lottery model in which two species are similar in many respects. Each species has equal death probabilities $\delta$, equal spatial variances  ${\sigma^{(t)}}^2$, equal temporal variances ${\sigma^{(t)}}^2$, and equal space-time interaction coefficients $\theta$. The two species only differ in how they respond to the environment (i.e., $\phi^{(x)} < 1$, $\phi^{(t)} < 1$). 

Various tricks can be used to simplify the expressions for the small-noise coexistence mechanisms. In the variance and covariance terms inherent the in small-noise coexistence mechanisms, the competition parameter can be expressed in terms of the environmental parameter, by 1) Taylor-series expanding competition with respect to the $E_j$ and $n_j$, 2) substituting into the covariance terms and truncating at first order in accordance with the small-noise assumptions, 3) recognizing that $\Cov{}{E_j}{n_j} = 0$ because the environment spatially and temporally uncorrelated, 4) recognizing that $\Var{}{n_r} = 0$ in the case of two species, since $n_r$ is fixed at 1. To compute fitness-density covariance, $\Delta \kappa$, we must first calculate the quasi-steady spatial distribution of the invader (see Section \ref{sec:Extended discussion of invasion growth rates, including computational tricks}). In Appendix \ref{app:Deriving the small-noise fitness-density covariance for the spatiotemporal lottery model}, we derive an approximation of this distribution using perturbation theory, recursion, and the geometric series.

\begin{tcolorbox}[breakable, title = Small-noise coexistence mechanisms in the spatiotemporal lottery model: two symmetric species with diffuse competition, subtitle style={boxrule=0.4pt,
colback=black!30!white}]

\tcbsubtitle{Density-independent effects}

\begin{equation} \label{dE_LM}
\begin{aligned}
 \Delta E_{i}  = 0
\end{aligned}
\end{equation}

\begin{flalign}
     \Delta E_{i,A} & = 0 & \\
     \Delta E_{i,S} & = 0 & \\
     \Delta E_{i,T} & = 0  & \\
     \Delta E{i,R} & = 0 & 
\end{flalign}

\tcbsubtitle{Linear density-dependent effects}
      
\begin{flalign}
   \Delta\rho_i = 0
\end{flalign}      

\tcbsubtitle{Relative nonlinearity}
      
\begin{equation}
\begin{aligned}
\Delta N_{i} = 0
\end{aligned}
\end{equation}

\begin{flalign}
    \Delta N_{i,S} & = 0 & \\
    \Delta N_{i,T} & = 0 & \\
    \Delta N_{i,R} & = 0 & 
\end{flalign}

\tcbsubtitle{The storage effect}
      
\begin{equation}
\begin{aligned}
\Delta I_{i} = \delta_i \left[ {\sigma^{(x)}}^2 \left( 1 -  \phi_{ir}^{(x)} \right) + {\sigma^{(t)}}^2 \left[ \left( \delta_i - 1 \right) \phi_{ir}^{(x)} -  \left( \delta_r - 1 \right) \right]  + \theta^2 {\sigma^{(x)}}^2 {\sigma^{(t)}}^2 \left( 1 -  \phi_{ir}^{(x)} \phi_{ir}^{(t)} \right) \right]
\end{aligned}
\end{equation}
 
\begin{flalign}
    \Delta I_{i,A} & = 0 & \\
    \Delta I_{i,S} & =  \delta_i {\sigma^{(x)}}^2 \left( 1 -  \phi_{ir}^{(x)} \right) & \\
    \Delta I_{i,T} & = \delta_i {\sigma^{(t)}}^2 \left[ \left( \delta_i - 1 \right) \phi_{ir}^{(t)} -  \left( \delta_r - 1 \right) \right] \label{Eq:lottery model:temporal storage effect} & \\
    \Delta I_{i,R} & = \delta_i \theta^2 {\sigma^{(x)}}^2 {\sigma^{(t)}}^2 \left( 1 -  \phi_{ir}^{(x)} \phi_{ir}^{(t)} \right) & 
\end{flalign}

\tcbsubtitle{Fitness-density covariance}
      
\begin{flalign} \label{dkappa_LM} 
    \Delta\kappa_i = \frac{2 q \delta_i^2 {\sigma^{(x)}}^2}{1-q} \left[\theta^2 {\sigma^{(t)}}^2 \left(1 - \phi_{ir}^{(x)} \phi_{ir}^{(t)} \right) - \phi_{ir}^{(x)} \right]
\end{flalign}     
\end{tcolorbox}

\subsection{Discussion of coexistence in the spatiotemporal lottery model}
\label{subsec:Discussion of coexistence in the spatiotemporal lottery model}

We first use the small-noise coexistence mechanisms above to look at edge cases where there is no spatial or temporal variation. When there is no spatial variation (i.e., $\sigma^{(x)} = 0$, the lottery model analyzed in this section collapses to the temporal lottery model of \textcite{Chesson1994}. The entire invasion growth rate is $\delta \left(\sigma^{(t)}\right)^2 (1 - \delta) ( 1-\phi^{(t)})$, which transparently shows that stable coexistence is not possible if species' responses to the environment are perfectly correlated ($\phi^{(t)} = 1$), or if generations are non-overlapping ($\delta = 1$). This latter result speaks to the storage effect's namesake: coexistence "... relies on such buffering effects of persistent stages..." (\cite{Chesson2003}).

When there is no temporal variation and we assume no local retention (i.e., $\sigma^{(t)} = 0$ and $q = 1$), our lottery model collapses to the spatial lottery model of \textcite{chesson2000general}. In this case, the invasion growth rate is $\delta \left(\sigma^{(x)}\right)^2 ( 1-\phi^{(x)})$, which demonstrates that coexistence is possible in the face of non-overlapping generations (i.e., when $\delta = 1$).

Finally, we consider the spatiotemporal lottery model. The invasion growth rate, minus fitness-density covariance and any space-time interaction terms (i.e., $\Delta I_{i,R}$), is $\delta( \left(\sigma^{(t)}\right)^2 (1 - \delta) ( 1-\phi^{(t)}) + \left(\sigma^{(x)}\right)^2 ( 1-\phi^{(x)}))$, the sum of invasion growth rates in the purely-temporal-variation case and the only-spatial-variation case. This quantity shows us that while spatial and temporal variation both tend to promote coexistence, they do not do so symmetrically. Specifically, compared to spatial variation, temporal variation is discounted by a factor of $(1-\delta)$. This discrepancy can be explained by the tendency of temporal variation  to decrease the geometric mean of $\lambda_j$ (\cite{Lewontin1969}).
 
Next, consider the sum of all remainder terms from the space-time decomposition, which is equal to $\delta \theta^2 (1- \phi^{(x)} \phi^{(t)}) \left(\sigma^{(x)} \sigma^{(t)}\right)^2$. Both this quantity and the small-noise fitness density covariance (\eqref{dkappa_LM}) reveal that even when generations are overlapping and responses to time-effects are perfectly correlated across species (i.e., $\phi^{(t)} = 1$), temporal variation can still promote coexistence by effectively amplifying species-specific responses to spatial variation, with strength according to the interaction coefficient $\theta$. Note, however, that this result is a consequence of the artificial assumption that the interaction between space and time effects, $\theta$, is large. Also note that when both species respond identically to patch effect and time effects, the space-time interaction terms disappear, confirming the perennial fact that niche differences are required for stable coexistence.

Our analysis in the preceding paragraph reveals that sometimes, the components of the space-time decomposition of coexistence mechanisms are not of fundamental interest. One may wish to aggregate terms in various ways, e.g., all space terms, all remainder terms, all terms containing partial derivatives of $C$ (i.e., both $\Delta \rho_i$ and $\Delta N_i$). Conversely, the invasion growth rate partition can be made even more fine-grained. \textcite{Ellner2019} decomposed $\Delta E_{i}$ into multiple terms, and partitioned the invasion growth rates with respect to trait values (as opposed to $E$ and $C$). In many models, the competition parameter $C_j$ can be expressed as a function of multiple regulating factors (see Appendix \ref{app:Generalization of MCT to different classes of models:Continuous-time models:Multiple regulating factors}), so naturally, $\Delta \rho_i$, $\Delta N_i$, and $\Delta I_i$ can be broken down further into terms which measure the contributions of individual (or subsets of) regulating factors. 

\section{Discussion}
\label{sec:Discussion}

\newpage
\begin{table}
\caption{\label{tab:co_max} The maximum number of species that can coexist via various coexistence mechanisms, in a system with $L$ discrete resources, $M$ discrete environmental states, and $K$ discrete patches. In the column headings, \textit{spatial variation} and \textit{temporal variation} refer to variation in the environment, regulating factors, and relative density. The entries in this table were derived as follows: only one species will have the largest $\Delta E$, and in the absence of other influences on the per capita growth rates, this species' relative frequency will approach 1 over time. The entries for $\Delta \rho$ simply express the competitive exclusion principle. The entries for $\Delta N$ follow from recognizing that the covariances between regulating factors can be treated as honorary regulating factors, and then by applying the competitive exclusion principle. The entries for $\Delta I$ are derived in the same way, and are an obvious extrapolation of the work by \cite{miller2017evolutionary}. The entries for $\Delta \kappa$ come from Appendix \ref{app:The maximum number of species that can coexist via fitness density covariance }. It is also well known that many species can coexist if patches have different resource supply points (\cite{levins1974discussion}; \cite{tilman1982resourceST}; \cite{chase2003ecological}); this manifests as the $M \times L$ term in the entries for $\Delta \kappa$, where $M$ is the number of distinct resource supply points. We have formally analyzed the case where fitness-density covariance is caused by aggregating behavior (such as swarming or schooling) or preferential dispersal (\cite{barabas2018chesson}, Appendix S5), but we imagine that behaviors or patch preferences can be treated as density-independent variables, and therefore, that the table entries for $\Delta \kappa$ are still accurate. }

\begin{tabular}{ >{\raggedright}p{0.4 \linewidth} >{\centering\arraybackslash} p{0.1\linewidth} >{\centering\arraybackslash} p{0.2\linewidth} >{\centering\arraybackslash} p{0.15 \linewidth} >{\centering\arraybackslash} p{0.2 \linewidth} }

Coexistence mechanisms & Models with neither spatial nor temporal variation & Models with only spatial variation & Models with only temporal variation & Models with spatiotemporal variation \\
\toprule
$\Delta E$: Density-independent effects & 1 & 1 & 1 & 1 \\
$\Delta \rho$: Linear, density-dependent effects & L & $L$ & $L$ & $L$ \\
$\Delta N$: Relative nonlinearity & 0 & $(L(L-1))/2$ & $(L(L-1))/2$ & $L(L-1)$ \\
$\Delta I$: Storage effect & 0 & $L M$ & $L M$ & $2 L M$ \\
$\Delta \kappa$: Fitness-density covariance & 0 & $LM +(L(L-1))/2$ & 0 & $LM +(L(L-1))/2$ \\
\bottomrule
\end{tabular}
\label{tab:max species}
\end{table}

In this paper, we have shown how the invasion growth rate can be partitioned so as to isolate the effects of spatial variation and temporal variation. With this new capability, one can determine whether species are coexisting because of spatial heterogeneity, temporally changing environmental conditions, or both. Further, one can break-down individual coexistence mechanisms (such as the storage effect) in to contributions from spatial and temporal variation, e.g., the spatial storage effect and the temporal storage effect can be extracted from a complex model with spatiotemporal variation.

To calculate the invasion growth rate in spatiotemporal models, one must average local growth rates over space and time. However, a simple arithmetic average over space and time is not appropriate, due to a fundamental difference in how populations grow over space and time: with respect to the geometric mean of the finite rate of increase (the quantity predictive of persistence; \cite{Metz1992}), contributions from populations across space are additive, but contributions from populations across time are multiplicative. Therefore, the appropriate spatiotemporal averaging (given by \eqref{time_decomp_2}) involves a density-weighted spatial average, followed by a temporal average on the log-scale.

To isolate the effects of spatial and temporal variation, we first define a reference state where both spatial and temporal variation are turned off; then, we separately turn on spatial (temporal) variation, and identify the difference as the main effect of spatial (temporal) variation. Put in such colloquial terms, this procedure may appear \textit{ad hoc} at first glance. However, we show that this procedure agrees with intuition in a simple example (Appendix \ref{app:Justification of the space-time decomposition:A toy model with only spatially or only temporally varying abiotic factor}), and is concordant with philosophical accounts of causation (Appendix \ref{app:Justification of the space-time decomposition:The space-time decomposition measures causation}). In statistics (e.g., multivariate regressions, ANOVA, directed acyclic graphs), the term \textit{effect of X} is often used to describe the marginal effects of X is relation to some reference state (\cite{vanderweele2015explanation}). Therefore, our space-time decomposition is a natural extension of ordinary scientific practice. 

A few basic insights emerge from Spatiotemporal Modern Coexistence Theory (MCT). The inclusion of spatiotemporal fluctuations (as opposed to only spatial or only temporal fluctuations) exactly doubles the maximum number of species that the fluctuation-dependent coexistence mechanisms can support (Table \ref{tab:max species}). The reason is laid bare in the space-time decomposition of the small-noise coexistence mechanisms (\eqref{dr EA}--\eqref{dIR}): species may specialize on either spatial variation \textit{or} temporal variation. It is worth noting that this result depends on the veracity of the small-noise assumptions (Appendix \ref{app:Deriving small-noise coexistence mechanisms:Small noise assumptions}); even more species could potentially coexist by specializing on higher-order moments (\cite{zicarelli1975mathematical}; \cite{levins1979coexistenceST}), such as the spatial skew of resource concentrations. 

Table \ref{tab:max species} reveals that with even a modest number of regulating factors and environmental states, there are more than enough ways for species to coexistence. This highlights the importance of actually measuring coexistence mechanisms in real communities. Table \ref{tab:max species} also shows the enormous potential of the fluctuation-dependent coexistence mechanisms, relative to classical explanations for coexistence (i.e., $\Delta \rho_i$). While this may be interesting, it is not likely to drive diversity patterns in the real world. For one, it has been argued that regulating factors are plentiful if you look hard enough (\cite{levin1970community}; \cite{haigh1972can}; \cite{abrams1988should}). Second, biodiversity is affected by many forces, including structural stability (\cite{Gyllenberg2005}), evolutionary / developmental / physiological constraints on extreme forms of specialization, and extinction--speciation balance.

Spatiotemporal MCT also strengthens an \textit{a priori} refutation of the \textit{competitive exclusion principle} (the idea that no more than $L$ species can coexist on $L$ regulating factors). The competitive exclusion principle was originally based on equilibrium theory, but the principle still applies in fluctuating environments when there are no fluctuation-dependent coexistence mechanisms (\cite{hening2020competitive}; \cite{barabas2018chesson}, p.295). Of course, for this to occur, there must be linear responses to regulating factors (this precludes relative nonlinearity) and no interaction effect between environment and competition (this precludes the storage effect).  Spatiotemporal MCT shows that species' responses to regulating factors cannot simultaneously be linear with respect to fluctuations on the natural scale (i.e., $\pdv[2]{g_j}{C_j} = \beta_j^{(2)} = 0$), which is necessary for spatial averaging, and linear with respect to fluctuations on the log-scale (i.e., $\pdv[2]{\log(g_j)}{C_j} = \beta_j^{(2)} - \beta_j^{(1)^2}  = 0$), which is necessary for temporal averaging. This shows that the competitive exclusion principle is unlikely to be applicable in the real world, even if we could count the number of regulating factors. 

Though spatiotemporal MCT has produced some theoretical insights, its primary value is as a methodology for inferring the mechanisms of coexistence in real communities. Spatiotemporal MCT allows for the analysis of more realistic models, which naturally lead to better inferences. Although generating realistic models requires immense amounts of system-specific knowledge, data collection, and statistical expertise, all of this hard work can be thought of as a safeguard against bad inferences. When simplistic statistical approaches are used to understand community structure, the data is often overdetermined by theory. For example, left skew in a species abundance distributions could indicate neutral population dynamics (\cite{hubbell2001unified}); or temporal autocorrelation in sampling (\cite{mcgill2003does}); or an excess of transient species (\cite{Magurran2003}); or a sequential stick-breaking model (\cite{nee1991lifting}); or a log-normal distribution paired with a zero-sum constraint (\cite{pueyo2006diversity}). Randomization-based null models for detecting interspecific competition can implicitly exclude  or include the effects of competition (\cite{Connor1979}, \cite{Diamond1982}). A saturating curve on a plot of regional vs. local species richness could indicate environmental filtering (\cite{Cornell1992}) or dispersal limitation (\cite{Fox2000}).

While data is always overdetermined by theory to some extent (\cite{DuhemPierreMauriceMarie1954Taas}), the problem can be abated by MCT's \textit{model-based} approach and  a few \textit{best practices}. First, one ought to large / flexible / complex models. Such models are less biased, and implicitly capture structural uncertainty (\cite{draper1995assessment}) in the form of parameter uncertainty (e.g. for the student's \textit{t}-distribution, the \textit{degrees of freedom} parameter interpolates between a gaussian distribution and and a cauchy distribution). As Leonard Savage used to say, all models should be "as big as a house" (qtd in \cite{draper1995assessment}). Simple "template models" (like the annual plant model; \cite{lanuza2018opposing}) can be made complex through the process of \textit{continuous, iterative model expansion} (\cite{box1980sampling}; \cite{draper1995assessment}; \cite{gelman2020bayesian}; \cite{gelman2020bayesian}). 

Another modelling "best practice" is to propagate uncertainty in model parameters through to the level of coexistence mechanisms, which can be generically accomplished by sampling from bootstrap or posterior distributions of model parameters. To our knowledge, only one empirical application of MCT (\cite{ellner2016quantify}, Section SI.8) has performed this crucial step. Without uncertainty propagation, it is difficult to say whether estimates of coexistence mechanisms reflect reality or sampling error.

Although we have extended MCT to more complex models, there remain a number of problems with MCT, primarily concerning the external validity of invasion growth rates as a measure of coexistence. But we should not be surprised nor disheartened that such problems exist: MCT was invented to explain the role of environmental variation in coexistence (\cite[p.~288]{barabas2018chesson}, \cite[p.~6]{Chesson2019}), not to be a methodology. There has been a recent surge of interest in the interpretation and application of MCT (\cite{ellner2016quantify}; \cite{Ellner2019}; \cite{Grainger2019}; \cite{Song2020}; \cite{pande2020mean}; \cite{Ellner2020}; \cite{barabas2018chesson}; \cite{Chesson2019}; and \cite{barabas2020chesson}; \cite{johnson2022methods}; \cite{johnson2022storage}; \cite{johnson2022towards}), but more work needs to be done.

\section{Acknowledgements}
We would like to thank Simon Stump and Sebastian Schreiber for discussions; and Logan Brissette for copy editing. This research is supported in part by NSF Grant DMS – 1817124 Metacommunity Dynamics: Integrating Local Dynamics, Stochasticity and Connectivity.

\section{Appendixes}
\label{app}
 
\subsection{Deriving small-noise coexistence mechanisms} 
\label{app:Deriving small-noise coexistence mechanisms}

The derivation of spatiotemporal coexistence mechanisms can be broken into four parts. In part 1, the local finite rate of increase is expressed in a common format: a polynomial of $E_j$ and $C_j$. This requires expressing a model of population dynamics as function of $E_j$ and $C_j$ (Section \ref{app:Deriving small-noise coexistence mechanisms:Population growth as a function of the environment and competition}), assuming that environmental fluctuations are small (Section \ref{app:Deriving small-noise coexistence mechanisms:Small noise assumptions}), and applying a Taylor series expansion (Section \ref{app:Deriving small-noise coexistence mechanisms:Decomposing the finite rate of increase: The quadratic approximation}). In part 2, the appropriate spatial (Section  \ref{app:Deriving small-noise coexistence mechanisms:Spatial averaging and fitness-density covariance}) and temporal (Section \ref{app:Deriving small-noise coexistence mechanisms:Temporal averaging}) averaging is applied in order to express the invasion growth rate in terms of local finite rates of increase. In part 3, The approximations derived in part 1 and 2 are combined to create a long expression for each species' average growth rate (Section \ref{app:Putting it all together: A decomposition of the average growth rate}). In part 4, the small-noise coexistence mechanisms are finally produced (formulas presented in the main text) by comparing the invader to the residents.

Why does our exposition feature discrete-time populations dynamics? For one, the connection with data-based modelling of real communities is more transparent, since data is collected at discrete points in time, and as a consequence, ecologists primarily fit discrete-time models. Secondly, the expressions for the small-noise coexistence mechanisms in the case of discrete-time are identical to those in case of continuous-time when environmental stochasticity is proportional to white noise (Section \ref{app:Generalization of MCT to different classes of models:Continuous-time models}).

A brief technical note: Throughout the paper, we use the notation $E_j$ as shorthand for $E_j(x,t)$; it is \textit{not} the case that that $E_j$ is a random variable and that $E_j(x,t)$ is a realization of said random variable, as the notation seems to imply. As \textcite{chesson2000general} points out, the notation can be made more precise by adding the seed number/ sample path as an additional argument, such that $E_j(x,t,\omega)$ is a realization of the random variable $E_j(x,t)$. Throughout this paper, when we apply the expectation operator (or covariance or variance operators), we sum over space and/or time while fixing the sample path $\omega$.

\subsubsection{Population growth as a function of the environment and competition}
\label{app:Deriving small-noise coexistence mechanisms:Population growth as a function of the environment and competition}

The local finite rate of increase, $\lambda_j$, is given by the function $g_j(E_j, C_j)$, where $E_j$ represents the effects of density-independent factors and $C_j$ represents the effects of density-dependent factors (also known as regulating factors or limiting factors). 

The parameter $E_j$ has many names: the \textit{environmentally-dependent parameter}, the \textit{response to the environment}, the \textit{environmental parameter}, or the \textit{environment}. It is typically a demographic parameter that depends on the abiotic environment, such as per capita fecundity or the probability of seed germination, hence the terminology \textit{response to the environment}. But $E_j$ may also be a literal environmental variable, such as annual precipitation, degree days, or soil type. It is important to keep in mind that $E_j$ need not represent the effects of the abiotic environment, since not all density-independent factors are part of the abiotic environment (e.g., mortality from a generalist predator), and not all density-dependent factors are biotic (e.g., refugia, soil nutrients).

The parameter $C_j$ is often called the \textit{competition parameter}, or simply \textit{competition}. Concrete examples of the competition parameter are the number of juvenile fish competing per open territory in the lottery model, or a linear combination of population densities, as in the competitive Lotka-Volterra model. The focus on competition reflects MCT's intellectual origin (and more generally, ecology's bias towards competition; \cite[p.~164]{MittelbachGaryGeorge2019Ce}) but the density-dependent $C$ can just as easily represent predation pressure (\cite{kuang2010interacting}; \cite{chesson2010storage}; \cite{stump2015distance}; \cite{stump2017optimally}) or mutualistic benefits (\cite{stump2018spatial}). 

We note that in some papers (e.g., (\cite{Chesson1994}; \cite{Chesson2018}; \cite{ellner2016quantify}), $C_{j}^{\{-i \}}$ or $C_{j\backslash i}$ is used to denote the competition parameter of species $j$ when species $i$ is absent. We simply use $C_j$ to denote the same, since we are always considering a community in which one species is the invader. 

\subsubsection{Decomposing the finite rate of increase: the quadratic approximation}
\label{app:Deriving small-noise coexistence mechanisms:Decomposing the finite rate of increase: The quadratic approximation}

We will decompose $g_j(E_j,C_j)$ via a second-order Taylor series expansion. First though, we must select \textit{equilibrium values} of the environment and competition to expand about. These values, denoted $E_j^*$ and $C_j^*$, must be selected so that that $g_j(E_j^*,C_j^*) = 1$, which functions to eliminate the zeroth-order Taylor series coefficient (see \eqref{taylor_decomp}). 

In general, there is no unique choice of $E_j^*$ and $C_j^*$, though as \textcite{Chesson1994} notes, fixing one parameter will determine the other. That being said, not all choices are equally appropriate. In particular, for every term in the Taylor series expansion to be the same order of magnitude -- and thus of commensurate importance -- we must simultaneously select $E_j^*$ to be close to $\E{x,t}{E_j}$, and $C_j^*$ to be close to $\E{x,t}{C_j}$ (the reasoning will be explained in the following section; \ref{app:Deriving small-noise coexistence mechanisms:Small noise assumptions}). 

There is a canonical method for selecting $E_j^*$ and $C_j^*$: virtually eliminate environmental noise, select $E_j^*$ as the environmental parameter in the resulting \textit{deterministic skeleton} (\cite{Coulson2004}), and then select $C_j^*$ based on the constraint $g_j(E_j^*,C_j^*) = 1$. In many models, the species-specific competition parameter $C_j$ can be expressed as a species-specific function of shared regulating factors (e.g., species densities, mineral nutrients). In such models, if one desires to quantify the contributions of individual regulating factors to various coexistence mechanisms (see Section \ref{app:Generalization of MCT to different classes of models:Continuous-time models:Multiple regulating factors}), then one must select equilibrium values of these regulating factors.
When there are multiple regulating factors, there are an infinite number of ways to select their equilibrium values -- there are multiple unknowns and just one constraint ($g_j(E_j^*,C_j^*) = 1$) -- but there are several reasonable strategies (see \cite{johnson2022methods}, Section 2.1).

With the appropriate selection of the equilibrium values, we expand the local finite rate of increase with a second-order Taylor Series about $E_{j}^{*}$ and $C_{j}^{*}$:
\begin{equation} \label{taylor_decomp}
\begin{aligned}
 \left. {g_{j}(E_j,C_j)}%
_{\stackunder[1pt]{}{}}%
 \right|_{%
 \stackon[1pt]{$\scriptscriptstyle E_j = E_{j}^{*}$}{$\scriptscriptstyle C_j = C_{j}^{*}$}}
 \approx \; & 1 + \alpha_j^{(1)} (E_j - E_{j}^{*}) + \beta_j^{(1)} (C_j - C_{j}^{*}) \\ & + 
\frac{1}{2} \alpha_j^{(2)} (E_j - E_{j}^{*})^{2} + \frac{1}{2} \beta_j^{(2)} (C_j - C_{j}^{*})^{2} + 
\zeta_j  (E_j - E_{j}^{*})   (C_j - C_{j}^{*}).
\end{aligned}
\end{equation}
The coefficients of the Taylor series are
\begin{equation}  \label{taylor_coef}
\begin{aligned}
 \alpha_j^{(1)} = \pdv{g_j\scriptstyle{(E_j^*, C_j^*)}}{E_j},  \quad
 \beta_j^{(1)} = \pdv{g_j\scriptstyle{(E_j^*, C_j^*)}}{C_j},  \quad
 \alpha_j^{(2)} = \pdv[2]{g_j\scriptstyle{(E_j^*, C_j^*)}}{E_j},  \quad
 \beta_j^{(2)} = \pdv[2]{g_j\scriptstyle{(E_j^*, C_j^*)}}{C_j,}  \quad
 \zeta_j = \pdv{g_j\scriptstyle{(E_j^*, C_j^*)}}{E_j}{C_j}.
\end{aligned}
\end{equation}

\subsubsection{Small noise assumptions}
\label{app:Deriving small-noise coexistence mechanisms:Small noise assumptions}

In order for the second-order Taylor series expansion (the r.h.s. of \eqref{taylor_decomp}) to be a good approximation of $g_j(E_j, C_j)$, we must make some assumptions about the magnitude of environmental fluctuations. First, we assume that the environmental parameter $E_j$ fluctuates about $E_j^*$ in a small finite range, and that the size of this range in controlled by a small parameter $\sigma$. Here, we use the conventional "big-oh" notation to denote an upper bound on magnitude of fluctuations:
\begin{equation}
	 E_j - E_j^* = \mathcal{O}(\sigma).  \label{ass:env_fluc}
\end{equation}
More precisely, this means that $\abs{E_j - E_j^*} < k \sigma$, with some constant $k$ as $\sigma \rightarrow 0$. Our next assumption states that environmental fluctuations are even smaller when averaged across space and time:
\begin{equation}
	 \E{x,t}{E_j} - E_j^* = \mathcal{O}(\sigma^2).  \label{ass:avg_env_fluc}
\end{equation}
Note that the spatiotemporal average of fluctuations is much smaller than maximum fluctuation, since the square of small number is much smaller than that number. The justification of the above assumption is either 1) that positive and negative fluctuations cancel out, or 2) that large fluctuations (which set the magnitude of $E_j - E_j^*$) are overpowered by many smaller fluctuations. Functionally, the assumption ensures that the effects of spatiotemporal averages are on the same order of magnitude as the effects of spatiotemporal variance (note that \eqref{ass:env_fluc} and \eqref{ass:avg_env_fluc} imply that $\Var{x,t}{E} = \mathcal{O}(\sigma^2)$ ).

To help make sense of the above assumptions, consider an environmental parameter $E_j(x,t) = a(x) + b(t)$. Both the patch effect $a(x)$ and time effect $b(t)$ independently take the value $+\sigma$ or $-\sigma$ with probability $ = 0.5$. By construction, the first assumption \eqref{ass:env_fluc} is met. If we then select $E_j^* = 0$, the relevant bounds are $\abs{E_j - E_j^*} \leq 2\sigma$, $\abs{\E{t}{E_j} - E_j^*} \leq \sigma$, and $\abs{\E{t}{E_j} - E_j^*} \leq \sigma$. Here we see that spatial and temporal averages of environmental fluctuations are on the same order of magnitude as the raw fluctuations, $E_j - E_j^*$. Furthermore, we see that $\E{x,t}{E_j} - E_j^* = 0$, which neatly demonstrates that the spatiotemporal average of fluctuations is exceedingly small (\eqref{ass:avg_env_fluc}).

Throughout this paper, we will refer to the two assumptions (\eqref{ass:env_fluc} and \eqref{ass:avg_env_fluc}) as \textit{small-noise assumptions}. Under fairly unrestrictive conditions, the small-noise assumptions can be used to prove analogous bounds for the competition parameter ($ C_j - C_j^* = \mathcal{O}(\sigma)$ and $\E{x,t}{C_j} - C_j^* = \mathcal{O}(\sigma^2)$) and for relative density ($ \nu_j - 1 = \mathcal{O}(\sigma)$ and $\E{x,t}{\nu_j} - 1 = \mathcal{O}(\sigma^2)$). Heavily paraphrased, the conditions are 1) that competition is a function of population densities and environmental responses (\cite[p.~269]{Chesson1994}); 2) that competition does not amplify itself over time (\cite[p.~269]{Chesson1994})); and 3) that "... any increase in local density due to dispersal cannot increase competition any more than $\mathcal{O(\sigma)}$ above the maximum competition applicable if there were no dispersal." (\cite[p.~234]{chesson2000general}). For all the details, see Appendix 2 of \textcite{Chesson1994} and Appendix 3 of \textcite{chesson2000general}. The same small parameter, $\sigma$, controls fluctuations in all relevant quantities (i.e environment, competition, relative density), since all fluctuations are ultimately a product of fluctuations in the environment.
 
The small-noise assumptions serve two primary purposes. First, they allow us to truncate the Taylor series (\eqref{taylor_decomp}) at second order, thus limiting the number of coexistence mechanisms that we might simultaneously consider. Second, the small-noise assumptions allow us to use the \textit{small-noise approximation} for dynamical systems (\cite{gardiner1985handbook}), resulting in simple stochastic models that permit analytical expressions for important quantities, e.g., the covariance between environment and competition. See \cite{schreiber2021positively} for a worked example of the small-noise approximation in the context of coexistence theory.

As \textcite{Chesson1994} points out, the small-noise assumptions are not statements about the absolute magnitude of environmental fluctuations. Instead, they are statements about the magnitude of environmental fluctuations, relative to other demographic parameters in a particular model. This does not mean that fluctuation-dependent coexistence mechanisms are unimportant, since the deterministic dynamics (corresponding to comparatively large parameter values) produce small per capita growth rates near equilibrium. Because the invader is not near equilibrium, however, we must make the additional assumption that between-species differences in the effects of regulating factors / competition on per capita growth rates are $\mathcal{O}(\sigma^2)$ (see \cite[p.~238]{Chesson1994}).

When the small-noise assumptions (and the auxiliary conditions above) are not met, one can proceed with two risks. First, the small-noise coexistence mechanisms may not sum approximately to the invasion growth rate; they will "miss" important processes that promote or hinder coexistence. Second, the exact coexistence mechanisms may capture unknown processes that involve large environmental fluctuations, thus making the exact coexistence mechanism less interpretable. 

The small-noise assumptions above require large fluctuations to be impossible, not just improbable. Restricting fluctuations to a finite range ensures that growth rates will not be dominated by low-probability, high-impact events. The gain in internal validity comes at the cost of external validity: it is often reasonable to model the environmental response by a random variable with support on the positive real numbers. For example, recruitment in some marine animals appears to follow lognormal distributions (\cite{Hennemuth1980}; \cite{Ripley2006}). However, the exact coexistence mechanisms circumvent the finite range assumption entirely, as long as we exclude from consideration the unlikely scenario where the distributions of $E_j$ and $C_j$ are so fat-tailed that spatial, temporal, or spatiotemporal averages of $E_j$ and $C_j$ do not exist. Given the plethora of assumptions implicit in any ecological model, a violation of the finite range assumption is just one of many ways in which the results of an MCT analysis are provisional. 

\subsubsection{Spatial averaging and fitness-density covariance}
\label{app:Deriving small-noise coexistence mechanisms:Spatial averaging and fitness-density covariance}

Next, we will derive a decomposition of the metapopulation finite rate of increase, $\widetilde{\lambda}_j(t)$. Consider a community with $K$ distinct patches. The metapopulation finite-rate of increase can be calculated as simple average of each individual's finite rate of increase, or equivalently, a weighted average of each patch's finite rate of increase, with weights equal to the relative density of the population in that patch. To see the logic of the latter scheme, first note that
\begin{equation}
\widetilde{\lambda}_j(t) = \frac{\sum\limits_{x=1}^K n_j(x,t+1)}{\sum\limits_{x=1}^K n_j(x,t)}
= \frac{\sum\limits_{x=1}^K n_j(x,t+1)}{ K \E{x}{n_j(t)}}.
\end{equation}
Using the local dynamics (\eqref{local_lambda}) to substitute for $n_j(x,t+1)$ , we find that

\begin{equation} \label{metapop_lambda}
\widetilde{\lambda}_j(t)
= \frac{1}{K \; \E{x}{n_j(t)}} \; \sum\limits_{x=1}^K \left( n_j(x,t) g_j(E_j(x,t), C_j(x,t)) \right) + \frac{1}{K \; \E{x}{n_j(t)}} \; \sum\limits_{x=1}^K \left( m_j(x,t) - e_j(x,t) \right).
\end{equation}

To simplify the above expression, we would like second additive term (the spatial sum of net dispersal) to vanish. This can be accomplished by assuming either 1) that the system is "closed", i.e., no individuals can enter or leave the system of patches, or 2) that the community receives roughly as many immigrants as it loses emigrants. Scenario 1 is likely to be approximately true for communities that span entire ecosystems, or for communities with very specific habitat requirements (e.g., Californian plants endemic to serpentine soils; \cite{Harrison2006}). In either case, there is no immigration into the metacommunity, and emmigration out of the metacommunity results in mortality that can be treated as part of the local dynamics of marginal patches. Scenario 2 is likely to be approximately true when the habitat surrounding the focal area is similar enough to the habitat within the focal area, so that immigration and emigration balances out over the margin of the focal area. In other words, the focal area (which itself is not closed) is representative of a larger metacommunity which is effectively closed.  

Assuming that dispersal is negligible at the spatial scale of the metapopulation, and rearranging terms, \eqref{metapop_lambda} simplifies significantly,
\begin{equation}
\widetilde{\lambda}_j(t) = \E{x}{\frac{n_j(t)}{\E{x}{n_j(t)}} g_j(E_j(x,t), C_j(x,t))},
\end{equation}
thus revealing that the metapopulation finite rate of increase is a density-weighted average of local finite rates of increase. $\widetilde{\lambda}_j(t)$ can be decomposed further with the law of total covariance: 
\begin{equation} \label{metapop_decomp}
\begin{aligned}
\widetilde{\lambda}_j(t) & = \E{x}{\frac{n_j(t)}{\E{x}{n_j(t)}}} \E{x}{g_j(E_j(x,t), C_j(x,t))} + \Cov{x}{\frac{n_j(t)}{\E{x}{n_j(x,t)}}}{g_j(E_j(x,t), C_j(x,t))} \\ 
& = \E{x}{g_j(E_j(x,t), C_j(x,t))} + \Cov{x}{\nu_j(t)}{g_j(E_j(t), C_j(t))}   
\end{aligned}
\end{equation}
where $\nu_j$ is the \textit{relative density} of species $j$, defined precisely as $\nu_j(x,t)  = \frac{n_j(x,t)}{\E{x}{n_j(x,t)}}$.

The first term in \eqref{metapop_decomp} is the spatial average of local per capita growth rates. It will be decomposed further with the Taylor series decomposition (Section \ref{app:Deriving small-noise coexistence mechanisms:Decomposing the finite rate of increase: The quadratic approximation}). The second term is the covariance between relative-density and growth rates, which captures the ability of species $j$ to end up in locations where it has high fitness, though the mechanism is completely unspecified. This term is the precursor to \textit{fitness-density covariance}. 

\subsubsection{Temporal averaging}
\label{app:Deriving small-noise coexistence mechanisms:Temporal averaging}

The quantity which is predictive of persistence is not $\E{t}{\widetilde{\lambda}_k}$, but rather $\E{t}{\log \widetilde{\lambda}_j}$. The logarithmic transformation converts a product of $\widetilde{\lambda}_j$ into a sum of $\log(\widetilde{\lambda}_j)$, which facilitates average-taking. 

Conditions on the magnitude of fluctuations in $E_j$, $C_j$, and $\nu_j$ (Section \ref{app:Deriving small-noise coexistence mechanisms:Small noise assumptions}) can be used to show that $\widetilde{\lambda}_j = 1 + \mathcal{O}(\sigma)$ and $\E{t}{\widetilde{\lambda}_j} = 1 + \mathcal{O}(\sigma^2)$. The logarithm can now be decomposed with a Taylor series expansion

\begin{equation} \label{temp_avg_1}
\begin{aligned}
 \left. {\log(\widetilde{\lambda}_j)}%
_{\stackunder[1pt]{}{}}%
 \right|_{%
 \stackon[1pt]{$\scriptscriptstyle \widetilde{\lambda}_j = 1$}{}}
 \approx \widetilde{\lambda}_j - 1 -  \frac{1}{2} \left(\widetilde{\lambda}_j - 1 \right)^2.
\end{aligned}
\end{equation}

Utilizing the fact that $\E{t}{\left(\widetilde{\lambda}_j - 1 \right)^2} = \Var{t}{\widetilde{\lambda}_j} + \mathcal{O}(\sigma^4)$, we take the average over time to obtain the average growth rate:

\begin{equation} \label{time_decomp}
   \E{t}{\log(\widetilde{\lambda}_j)} \approx \E{t}{\widetilde{\lambda}_j} - 1 -  \frac{1}{2}  \Var{t}{\widetilde{\lambda}_j}. 
\end{equation}

Plugging the decomposition of $\widetilde{\lambda}_j(t)$ (\eqref{metapop_decomp}) into equation \eqref{time_decomp}, we find that the invasion growth rate can be approximated entirely in the moments of $\lambda_j$ and $\nu_j$.

\begin{equation} \label{time_decomp_1}
\begin{aligned}
   \E{t}{\log(\widetilde{\lambda}_j)} \approx \E{x,t}{\lambda_j} + \E{t}{\Cov{x}{\nu_j(t)}{\lambda_j}} - 1 -   \frac{1}{2} \Var{t}{\E{t}{\lambda_j}}.
\end{aligned}
\end{equation}

\subsubsection{Putting it all together: A decomposition of the average growth rate}
\label{app:Putting it all together: A decomposition of the average growth rate}

The Taylor series decomposition of $g_j(E,C)$ (\eqref{taylor_decomp}) can be plugged into \eqref{time_decomp_1}, producing a fine-grained partition of species $j$'s average growth rate

\begin{equation} \label{big_decomp_2}
\begin{aligned}
\left. {\E{t}{\log(\widetilde{\lambda}_j)}}%
_{\stackunder[1pt]{}{}}%
\right|_{%
\stackon[1pt]{$\scriptscriptstyle E_j = E_{j}^{*}$}{$\scriptscriptstyle C_j = C_{j}^{*}$}}
\approx \; &  \alpha_j^{(1)} \E{x,t}{(E_j - E_{j}^{*})} + \beta_j^{(1)} \E{x,t}{(C_j - C_{j}^{*})} \\ 
+ & \; \frac{1}{2} \alpha_j^{(2)} \Var{x,t}{E_j} + \frac{1}{2} \beta_j^{(2)} \Var{x,t}{C_j} + \zeta_j  \Cov{x,t}{E_j}{C_j} \\ 
+ & \; \E{t}{\Cov{x}{\nu_j}{ \alpha_j^{(1)} (E_j - E_{j}^{*}) + \beta_j^{(1)} (C_j - C_{j}^{*})}} \\ 
- & \; \frac{1}{2} \alpha_j^{(1)^{2}} \Var{t}{\E{x}{E_j}} - \frac{1}{2} \beta_j^{(1)^{2}} \Var{t}{\E{x}{E_j}} - \alpha_j^{(1)}\beta_j^{(1)} \Cov{t}{\E{x}{E_j}}{\E{x}{C_j}}.
\end{aligned}
\end{equation}

The additive terms in equation \eqref{big_decomp_2}, can be thought of as a components of the average growth rate, each of which captures some "effect" on population growth. The components are not generally independent, which implies that the consequent coexistence mechanisms are not generally independent (\cite{Song2020}; \cite{kuang2010interacting}; \cite{Yuan2015}). For instance, in the spatiotemporal lottery model (Section \ref{sec:Example: the spatiotemporal lottery model} in the main text) a single parameter modulates all coexistence mechanisms. However, growth rate components may be conceptualized as distinct processes, just as ecology and evolution are interdependent but conceptually distinct. 

Note that the term $\E{x,t}{(E_j-E_j^*)(C_j-C_j^*)}$ has been replaced with $\Cov{x,t}{E_j}{C_j}$, since  $\Cov{x,t}{E_j}{C_j} = \E{x,t}{(E_j-E_j^*)(C_j-C_j^*)} + \mathcal{O}(\sigma^3)$ via the small-noise assumptions. Analogous replacements have been made for other variance and covariance terms in \eqref{big_decomp_2}. These replacements are not a necessary part of MCT, but they do shorten the mathematical expressions, which is aesthetically pleasing.

\subsection{Justification of the space-time decomposition}
\label{app:Justification of the space-time decomposition}

\subsubsection{A toy model with only spatially or only temporally varying abiotic factors}
\label{app:Justification of the space-time decomposition:A toy model with only spatially or only temporally varying abiotic factor}

Here, we aim to remove idle doubts about the appropriateness of our space-time decomposition by providing two justifications. Our first justification comes from the analysis of an edge case where the environmental response $E_j$ is a function of abiotic factors that individually vary only over space or time. This case is simple enough that we can describe our intuitions regarding what a space-time decomposition should do: The space component should only include the effects of the spatially varying abiotic factors; and the time component should only include the effects of the temporally varying abiotic factors.  
 
To be more concrete, consider two abiotic factors, $W$ and $Y$. $W$ only varies over space (i.e., at a particular location, $W$ does not vary from year-to-year) and $Y$ only varies over time (i.e., at a single point in time, all locations have the same value of $Y$).  Select the equilibrium values of the abiotic resources, $W^*$ and $Y^*$, so that $E_j^* = f_j(W^*, Y^*)$, where $f_j$ is a the function which relates abiotic factors to species $j$'s environmental response. The small-noise assumptions of MCT imply that $W - W^* = \mathcal{O}(\sigma)$, $Y - Y^* = \mathcal{O}(\sigma)$, $\E{x,t}{W - W^*} = \E{x}{W - W^*} =\mathcal{O}(\sigma^2)$, and $\E{x,t}{Y - Y^*} = \E{t}{Y - Y^*} = \mathcal{O}(\sigma^2)$. Using this information, we can derive expressions for the space-time decomposition of $\Var{x,t}{E_j}$. Applying a Taylor series of $f_j$ about $W^*$ and $Y^*$ for $E_j$, and utilizing the fact that the variance of a constant equals zero (e.g., $\Var{t}{W(x)} = 0$), the leading-order approximations of $S_j$, $T_j$, and $R_j$ (using \eqref{space term}--\eqref{remainder}) are

\begin{align}
& S_j \approx \left[ \pdv{f_j(W^*, Y^*)}{W} \right]^2 \Var{x}{W} \\
& T_j \approx \left[ \pdv{f_j(W^*, Y^*)}{Y} \right]^2 \Var{x}{W} \\ 
& R_j \approx \left[ \pdv[2]{f_j(W^*, Y^*)}{W}{Y} \right]^2 \Var{x}{W}\Var{t}{Y} \label{toy remainder}.
 \end{align}

The Taylor series coefficients show that $S_j$ captures the main effect of the spatially varying abiotic factor, $T_j$ captures the main effect of the temporally varying abiotic factor, and that $R_j$ captures the interaction effect between the two abiotic factors. This model is exceedingly simple, but it is the first line of evidence that our space-time decomposition behaves as desired. 

\subsubsection{The space-time decomposition measures causation}
\label{app:Justification of the space-time decomposition:The space-time decomposition measures causation}

Counterfactual theories of causation posit that causation can be explained in terms of counterfactual dependency (\cite{Hume1748}; \cite{mill1856system} , \cite{Lewis1973}, \cite{Pearl2018}). To say "A caused B", is to say "if A had not occurred, then B would not have occurred". To operationalize causation, we may calculate differences (with respect to some outcome of interest) between possible worlds, where the possible worlds are similar in every relevant way except for some focal causal factor. The comparison of possible worlds is crucial, which is why the counterfactual account of causation is sometimes called the difference-making account of causation. \textcite{Lewis1973} explains "We think of a cause as something that makes a difference, and the difference it makes must be a difference from what would have happened without it."  

The exposition above makes our challenge clear: to justify our space-time decomposition on the grounds that it captures causation, we must 1) describe $S_j$, $T_j$, and $R_j$ (see \eqref{space term} - \eqref{remainder}) in terms of differences between possible worlds, as has been done in the main text (Section \ref{sec:Spatiotemporal coexistence mechanisms:Exact coexistence mechanisms}) and 2) argue that the possible worlds in question are close in some relevant sense, following \posscite{Lewis1979} guideline that possible worlds "...maximize the spatiotemporal region thorough-out which perfect match of particular fact prevails". By using spatial (temporal) averaging to squash spatial (temporal) variation, we are doing just that: the sequence of spatial averages $A(t) = \E{x}{E_j}$ minimizes the squared error $\sum_{x,t} \left( E_j(x,t) -A(x) \right)^2$, under the constraints that there is no spatial variation, and that spatial variation must be squashed using only information from each individual time-step. 

\subsection{Deriving the small-noise fitness-density covariance for the spatiotemporal lottery model}
\label{app:Deriving the small-noise fitness-density covariance for the spatiotemporal lottery model}

The fitness-density covariance coexistence mechanisms (\eqref{dkappa}) is $\mathcal{O}(\sigma^2)$, which implies that $\Cov{x}{\nu_j(t)}{\lambda_j(t)} = \mathcal{O}(\sigma^2)$. Therefore, the leading-order approximation for the covariance terms will involve $\mathcal{O}(\sigma)$ approximations of $\nu_j(t)$ and $\lambda_j(t)$. To this end, we take a perturbative approach, expanding both parameters in powers of $\sigma$, $\nu_{j}(x,t) = \nu_{j,0}(x,t) + \sigma \nu_{j,1}(x,t) + ...$; and $\lambda_{j}(x,t) = \lambda_{j,0}(x,t) + \sigma \lambda_{j,1}(x,t) + ...$.

Matching like terms in the perturbative expansion and the Taylor series expansion of $\lambda_j$ (\eqref{taylor_decomp}), we find that $\lambda_{j,0}(x,t) = 1$ and $\sigma \lambda_{j,1}(x,t) = \alpha_j^{(1)} (E_j(x,t) - E_{j}^{*}) + \beta_j^{(1)} (C_j(x,t) - C_{j}^{*})$. The solution $\lambda_{j,0}(x,t) = 1$ implies that $\nu_{j,0}(x,t) = 1$. Noting the constancy of the zeroth-order solutions, the covariance can now be approximated as 

\begin{equation}
\begin{aligned}
 \Cov{x}{v_j(t)}{\lambda_j(t)} & = \Cov{x}{v_{j,0}(t) + \sigma v_{j,1}(t) + ...}{\lambda_{j,0}(t) + \sigma \lambda_{j,1}(t) + ...} \\
 & \approx \Cov{x}{\sigma v_{j,1}(t)}{\sigma \lambda_{j,1}(t)}.
\end{aligned}
\end{equation}

We now seek to simplify by expressing $v_{j,1}(x,t)$ in terms of the environmental parameter. Dividing both sides of the population map (\eqref{lottery}) by $\E{x}{N_j(t)}$, gives the relative-density map.

\begin{equation}
  \nu_j(x,t+1) =  q_j \nu_j(x,t) \frac{\lambda_j(x,t)}{\widetilde{\lambda}_j(t)} + 1 - q_j.
\end{equation}

The small-noise assumptions (Section \ref{app:Deriving small-noise coexistence mechanisms:Small noise assumptions}) allow us to make the substitution, $\widetilde{\lambda}_j(x,t) = 1 + \mathcal{O}(\sigma^2)$, which simplifies the relative density map to 

\begin{equation}
  \nu_j(x,t+1) =  q_j \nu_j(x,t) \lambda_j(x,t) + 1 - q_j.
\end{equation}

We now expand $v_j$ in powers of $\sigma$ and match terms of order $\sigma$. 

\begin{equation} 
  \mathcal{O}(\sigma): \quad \quad \nu_{j,1}(x,t+1) = q_j \nu_{j,1}(x,t) + q_j  \lambda_{j,1}(x,t).
\end{equation}

Substituting the above expression into the covariance produces

\begin{equation}
\begin{aligned}
\Cov{x}{v_j(t)}{\lambda_j(t)} & \approx \Cov{x}{\sigma v_{j,1}(t)}{\sigma \lambda_{j,1}(t)} \\
 = & \sigma^2 \Cov{x}{q v_{j,1}(t-1) + \lambda_j(t-1)}{\lambda_{j,1}(t)} \\
 = & \sigma^2 \Cov{x}{q^2 v_{j,1}(t-2) + q \lambda_{j,1}(t-2) + \lambda_{j,1}(t-1)}{\lambda_{j,1}(t)} \\
 \vdots \\
 = & \sigma^2 \sum \limits_{i=1}^{\infty} q^i \Cov{x}{\lambda_{j,1}(t-i)}{\lambda_{j,1}(t)}.
\end{aligned}
\end{equation}

Substituting  $\alpha_j^{(1)} (E_j(x,t) - E_{j}^{*}) + \beta_j^{(1)} (C_j(x,t) - C_{j}^{*}) + \mathcal{O}(\sigma^2)$ for $\sigma \lambda_{j,1}$, we get 

\begin{equation}
\begin{aligned} \label{fd_cov_ec}
 \Cov{x}{v_i(t)}{\lambda_i(t)}  \approx  \sum \limits_{s=1}^{\infty} q^s \mathrm{Cov}_x \bigl( & \alpha_i^{(1)} (E_i(x,t-i) - E_{i}^{*}) + \beta_i^{(1)} (C_i(x,t-s) - C_{i}^{*})  \\
 & \alpha_i^{(1)} (E_i(x,t) - E_{i}^{*}) + \beta_i^{(1)} (C_i(x,t) - C_{i}^{*}) \bigl).
 \end{aligned}
\end{equation}

Next, we express invader's competition parameter fluctuation in terms of the resident's environmental response. In the two-species lottery model, $C_i(x,t) - C_{i}^{*} = \frac{\partial C_i(E_r^*, N_r^*)}{\partial E_r}( E_r(x,t) - E_{r}^{*}) + \mathcal{O}(\sigma^2) =  E_r(x,t) - E_{r}^{*} + \mathcal{O}(\sigma^2)$. The covariance expression is now 

\begin{equation}
\begin{aligned}
 \Cov{x}{v_i(t)}{\lambda_i(t)}  \approx  \sum \limits_{s=1}^{\infty} q^s \mathrm{Cov}_x \bigl( & \alpha_i^{(1)} (E_i(x,t-i) - E_{i}^{*}) + \beta_i^{(1)} (E_r(x,t-s) - E_{r}^{*}) ,  \\
 & \alpha_i^{(1)} (E_i(x,t) - E_{i}^{*}) + \beta_i^{(1)} (E_r(x,t) - E_{r}^{*}) \bigl).
 \end{aligned}
\end{equation}

Finally, we write the environmental fluctuations in terms of patch and time effects (\eqref{lottery_noise}), evaluate the above expression using the geometric series and the symbols introduced in the Section \ref{sec:Example: the spatiotemporal lottery model}, e.g., $\Cov{x}{a_i}{a_r} = \phi_{ir}^{(x)} \sigma_{i}^{(x)} \sigma_{r}^{(x)}$, and take the average across time:

\begin{equation}
\begin{aligned}
 \E{t}{\Cov{x}{v_i}{\lambda_i}}  \approx  \frac{q}{1-q}\biggl[ & {\alpha_i^{(1)}}^2 {\sigma_i^{(x)}}^2 + {\beta_i^{(1)}}^2 {\sigma_r^{(x)}}^2 + 2\alpha_i^{(1)}\beta_i^{(1)} \phi_{ir}^{(x)} \sigma_i^{(x)}\sigma_r^{(x)} \\
 & + {\alpha_i^{(1)}}^2 \theta_i^2 {\sigma_i^{(x)}}^2 {\sigma_i^{(t)}}^2 + {\beta_i^{(1)}}^2 \theta_r^2 {\sigma_r^{(x)}}^2 {\sigma_r^{(t)}}^2 \\ 
 & + 2\alpha_i^{(1)}\beta_i^{(1)} \theta_i \theta_r \phi_{ir}^{(x)} \phi_{ir}^{(t)} \sigma_i^{(x)} \sigma_i^{(t)} \sigma_r^{(x)} \sigma_r^{(t)} \biggl].
 \label{FDCov complicated}
 \end{aligned}
\end{equation}

In the lottery model, there are always more larvae produced than are necessary to compensate for adult mortality. If there is only one resident, its densities will be exactly 1 everywhere after the local growth phase. Since global dispersal with local retention acts symmetrically on all patches, the resident's density will still be 1 everywhere after the dispersal phase. Therefore, the residents' covariance terms is zero, and the fitness density covariance coexistence mechanism is simply the expression above, \eqref{FDCov complicated}; i.e., $\Delta \kappa_i = \E{t}{\Cov{x}{v_i}{\lambda_i}}$. When symmetries in demographic parameters are taken into consideration, \eqref{FDCov complicated} reduces to the result in the main text, \eqref{dkappa_LM}.

\subsection{Generalization of MCT to different classes of models}
\label{app:Generalization of MCT to different classes of models}

\subsubsection{Continuous-time models}
\label{app:Generalization of MCT to different classes of models:Continuous-time models}

The continuous-time dynamics of a scalar population are given by the differential equations, 
\begin{equation} \label{local_r}
\frac{d n_j}{d t} = n_{j}(x,t) r_{j}(x,t) + c_j(x,t) - e_j(x,t) \qquad j = (1, 2, ..., S),
\end{equation}
where $r_j(x,t)$ is the instantaneous per capita growth rate or the intrinsic growth rate, and $c_j$ and $e_j$ are dispersal terms which will cancel out at the scale of the metapopulation (see Section \ref{app:Deriving small-noise coexistence mechanisms:Spatial averaging and fitness-density covariance}). For the most part, the derivation of coexistence mechanisms for continuous-time models follows the derivation in Section \ref{app:Deriving small-noise coexistence mechanisms}. However, the expressions for the single-species decomposition of the average growth rate (i.e., the continuous-time analogue of \eqref{big_decomp_2}) will depend on how stochasticity enters the population dynamics. 

First, define the metapopulation per capita growth rate as
\begin{equation}
    \widetilde{r}_j(t) = \E{x}{r(x,t)} + \Cov{x}{\nu_j(t)}{r_j(t)}, 
\end{equation}
in analogy with \eqref{metapop_decomp}. Next, consider the case where $E_j$ or $C_j$ fluctuates so rapidly that species $j$'s dynamics can be cast as a stochastic differential equation. The metapopulation density, $\widetilde{n}_j$, evolves according to
\begin{equation}
    d \widetilde{n}_j = \widetilde{n}_j \left[ \E{t}{\widetilde{r_j}} dt + \sqrt{\Var{t}{\widetilde{r_j}}} dW_j \right],
\end{equation}
where $dW_j$ is an increment of the Weiner Process, $\E{t}{\widetilde{r_j}}$ is the infinitesimal mean, and $\Var{t}{\widetilde{r_j}}$ is the infinitesimal variance (\cite{lande2003stochastic}; \cite{Braumann2007}). To calculate future population densities, and thus determine whether an invader will invade, the right-hand-side of the stochastic differential equation must be integrated across time. The two most popular calculi for evaluating stochastic integrals are Ito's calculus and Stratonovitch's calculus (\cite{RoughgardenJoan1979Topg}). 

\textcite{Braumann2007} convincingly showed that Ito's calculus is the correct choice when the infinitesimal mean is defined as the arithmetic (temporal mean of per capita growth rates. Under Ito's calculus, the solution to the stochastic differential equation is $\widetilde{n}_i(t) = \widetilde{n}_i(0) \exp{(\E{t}{\widetilde{r_i}} - \Var{t}{\widetilde{r_i}}/2) + \sqrt{\Var{t}{\widetilde{r_i}}} W(t)}$ (\textcite{Braumann2007}), so the invader only tends to increase when $(\E{t}{\widetilde{r_i}} - \Var{t}{\widetilde{r_i}}/2 > 0$. The discounting of the expected growth rate by half of the temporal variance should be reminiscent of \eqref{time_decomp}. In discrete time models, this discounting is an approximation that we justify using small-noise assumptions (Section \ref{app:Deriving small-noise coexistence mechanisms:Small noise assumptions}). Here, because stochastic differential equations are defined in the limit as the time-increment shrinks to zero, the noise is automatically small, and so the discounting is exact. 

To obtain expressions for the average growth rate that are analogous to \eqref{big_decomp_2}, we can expand $\E{t}{\widetilde{r_i}}$ and $\Var{t}{\widetilde{r_i}}$ with Taylor series; and truncate using the small-noise assumptions (Section \ref{app:Deriving small-noise coexistence mechanisms:Small noise assumptions}). Once expressions for average growth rates are in hand, the invader--resident comparison is straightforward. The Ito calculus produces
\begin{equation} 
\begin{aligned}
 \left. {\Biggl(\E{t}{\widetilde{r_j}} - \frac{\Var{t}{\widetilde{r_j}}}{2}\Biggr)}%
_{\stackunder[1pt]{}{}}%
 \right|_{%
 \stackon[1pt]{$\scriptscriptstyle E_j = E_{j}^{*}$}{$\scriptscriptstyle C_j = C_{j}^{*}$}}
 \approx \; &  \alpha_j^{(1)} \E{x,t}{(E_j - E_{j}^{*})} + \beta_j^{(1)} \E{x,t}{(C_j - C_{j}^{*})} \\ & + 
\frac{1}{2} \alpha_j^{(2)} \Var{x,t}{E_j} + \frac{1}{2} \beta_j^{(2)} \Var{x,t}{C_j} + 
\zeta_j  \Cov{x,t}{E_j}{C_j} \\ &
+ \E{t}{\Cov{x}{\nu_j}{ \alpha_j^{(1)} (E_j - E_{j}^{*}) + \beta_j^{(1)} (C_j - C_{j}^{*})}} \\ &
- \frac{1}{2} \alpha_j^{(1)^{2}} \Var{t}{\E{x}{E_j}} - \frac{1}{2} \beta_j^{(1)^{2}} \Var{t}{\E{x}{E_j}} - \alpha_j^{(1)}\beta_j^{(1)} \Cov{t}{\E{x}{E_j}}{\E{x}{C_j}},
\end{aligned}
\end{equation}
which is nearly identical to the discrete-time case (\eqref{big_decomp_2}), the only difference being that that the Taylor series coefficients are derivatives of $r_j$, not $\lambda_j$. For example, in the expression above,  $\alpha_j^{(1)} = \frac{\partial r_j(E_j^*, C_j^*)}{\partial E_j}$. 

When population densities are not governed by stochastic differential equations (regardless of whether $E_j$ or $C_j$ are, e.g., \textcite{li2016effects})  a simple arithmetic average over space and time gives the correct invasion growth rate:
 \begin{equation}
\begin{aligned}
 \left. {\E{t}{\widetilde{r_j}}}%
_{\stackunder[1pt]{}{}}%
 \right|_{%
 \stackon[1pt]{$\scriptscriptstyle E_j = E_{j}^{*}$}{$\scriptscriptstyle C_j = C_{j}^{*}$}}
 \approx \; &  \alpha_j^{(1)} \E{x,t}{(E_j - E_{j}^{*})} + \beta_j^{(1)} \E{x,t}{(C_j - C_{j}^{*})} \\ & + 
\frac{1}{2} \alpha_j^{(2)} \Var{x,t}{E_j} + \frac{1}{2} \beta_j^{(2)} \Var{x,t}{C_j} + 
\zeta_j  \Cov{x,t}{E_j}{C_j} \\ &
+ \E{t}{\Cov{x}{\nu_j}{ \alpha_j^{(1)} (E_j - E_{j}^{*}) + \beta_j^{(1)} (C_j - C_{j}^{*})}}.
\end{aligned}
\end{equation}
Here, there is no discounting for temporal variation, so spatial and temporal variation are treated symmetrically (with the exception of fitness density covariance). 

Finally, we note that it is often more difficult to fit a continuous-time model to data, since there are many trajectories that population densities can take between two successive observations. The typical way to fit such models is to convert a system of stochastic differential equations (or Langevin equations) into its 
Fokker-Planck representation, and then integrate the partial differential equation (for each observation) to get a probability density function (\cite{karlin1981second}; \cite{lande2003stochastic}).

\subsubsection{Multiple regulating factors}
\label{app:Generalization of MCT to different classes of models:Continuous-time models:Multiple regulating factors}

In the spatiotemporal lottery model, competition was a function of just one regulating factor: space. In more realistic models, we may want to cast competition as function of $L$ regulating factors, $\boldsymbol{F} = (F_1, F_2, ..., F_L)$. The regulating factors can be species densities, refugia, resources, or natural enemies, etc:

\begin{equation}
    C_j = \phi_j(\boldsymbol{F}).
\end{equation}

In previous work (e.g., \cite{Chesson1994}; \cite{barabas2020chesson}), the Taylor series coefficients $\beta_j^{(1)}$ and $\beta_j^{(2)}$ were absorbed into $\phi_{jk}^{(1)}$ and $\phi_{jkl}^{(2)}$. Our preferred notation makes the formulas slightly longer, but also implies that $\phi_j(\boldsymbol{F})$ can straightforwardly substituted for $C_j$, which in turn implies that the case of multiple regulating factors is not different from what is presented in the main text.  

With this approach, it is sometimes the case that $C_j = \lambda_j$, such that $\beta_j^{(1)} = \beta_j^{(2)} = 1$. This is similar to the approach taken by \textcite{Chesson2019}, who directly expands the growth rate with respect to the regulating factors by first defining $\phi'_j$ via the relationship $\mathscr{C}_j = \phi'_j(\boldsymbol{F})$ (where $\mathscr{C}_j = \log( \widetilde{\lambda}_j)$), and then expanding $\phi'_j$ with respect to the regulating factors.

\paragraph{Small-noise coexistence mechanisms}

Formulas for small-noise coexistence mechanisms which explicitly use the regulating factors can be obtained by taking the formulas for small-noise coexistence mechanisms in the main text (\eqref{dE} - \eqref{dkappa}), substituting in the Taylor series expansion of $\phi_j$ for $C_j$, and truncating using the small-noise assumptions. To taylor expand $\phi_j$, one must select equilibrium values $\boldsymbol{F^*}$ such that $C_j^* = \phi_j(\boldsymbol{F}^*)$. This task may be guided by the the requirements that $F_k - F_k^* = \mathcal{O}(\sigma)$ and  $\E{x,t}{F_k} - F_k^* = \mathcal{O}(\sigma^2)$, which are implied by the small-noise assumptions (Section \ref{app:Deriving small-noise coexistence mechanisms:Small noise assumptions}).

To keep the number of coexistence mechanisms from exploding beyond comprehension, all linear terms in $F_k$ get shunted into $\Delta \rho_i$, all nonlinear terms in $F_k$ get shunted into $\Delta N_i$, and all nonadditive terms in $F_k$ get shunted into $\Delta I_i$. This shunting helps to simplify and standardize, but users of MCT can still look at the relative roles of different regulating factors in promoting coexistence. We will present the small-noise coexistence mechanisms for a system with $L$ regulating factors, but first we introduce new notation to keep the expressions as simple as possible. Let $\phi_{jk}^{(1)} = \frac{\partial C_j}{\partial F_k}$ and  $\phi_{jkl}^{(2)} = \frac{\partial^2 C_j}{\partial F_k \partial F_l}$, where both quantities are evaluated at $E_j = E_j^*$ and $\boldsymbol{F} = \boldsymbol{F^*}$.

\begin{tcolorbox}[breakable, title = Formulas for small-noise coexistence mechanisms: multiple regulating factors, subtitle style={boxrule=0.4pt, colback=black!30!white}]

\tcbsubtitle{The invasion growth rate}

\begin{equation}
    \E{t}{\log(\widetilde{\lambda}_{i})} \approx \Delta E_{i} + \Delta\rho_i + \Delta\mathrm{N}_i + \Delta\mathrm{I}_i + \Delta\kappa_i,
\end{equation}

\tcbsubtitle{Density-independent effects}

\begin{flalign}
    \Delta E_{i} = & \left[\alpha_{i}^{(1)} \E{x,t}{(E_i - E_{i}^{*})} + \frac{1}{2} \alpha_i^{(2)}\Var{x,t}{E_i} - \frac{1}{2} \alpha_i^{(1)^{2}} \Var{t}{\E{x}{E_i}}\right] \nonumber \\ & 
    - \frac{1}{S-1} \sum\limits_{r \neq i}^S \frac{a_i}{a_r} \left[\alpha_r^{(1)} \E{x,t}{(E_r - E_{r}^{*})} + \frac{1}{2} \alpha_r^{(2)}\Var{x,t}{E_r} - \frac{1}{2} \alpha_r^{(1)^{2}} \Var{t}{\E{x}{E_r}}\right] 
\end{flalign}

\tcbsubtitle{Linear density-dependent effects}
      
\begin{flalign}
   \Delta\rho_i = \left[ \sum \limits_{k = 1}^L \beta_i^{(1)} \phi_{ik}^{(1)} \E{x,t}{F_k - F_k^{*}} \right] - \frac{1}{S-1} \left[ \sum\limits_{r \neq i}^S \sum \limits_{k = 1}^L \frac{a_i}{a_r} \beta_r^{(1)} \phi_{rk}^{(1)} \E{x,t}{F_k - F_k^{*}} \right]
\end{flalign}

\tcbsubtitle{Relative nonlinearity}
      
\begin{flalign}
     \Delta\mathrm{N}_i = & \frac{1}{2}  \left[\sum \limits_{k = 1}^L \sum \limits_{l = 1}^L \left(\beta_i^{(2)} \phi_{ikl}^{(2)} \Cov{x,t}{F_k}{F_l} - {\beta_i^{(1)}}^2 \phi_{ik}^{(1)} \phi_{il}^{(1)} \Cov{t}{\E{x}{F_k}}{\E{x}{F_l}} \right) \right] \nonumber \\
    & - \frac{1}{2(S-1)} \sum\limits_{r \neq i}^S \frac{a_i}{a_r} \left[ \sum \limits_{k = 1}^L \sum \limits_{l = 1}^L \left(\beta_i^{(2)} \phi_{rkl}^{(2)} \Cov{x,t}{F_k}{F_l} - {\beta_i^{(1)}}^2 \phi_{rk}^{(1)} \phi_{rl}^{(1)} \Cov{t}{\E{x}{F_k}}{\E{x}{F_l}} \right) \right]
\end{flalign}

\tcbsubtitle{The storage effect}
      
\begin{flalign}
     \Delta\mathrm{I}_i =  & \left[ \sum \limits_{k = 1}^L \left( \zeta_{i} \phi_{ik}^{(1)}  \Cov{x,t}{E_i}{F_k} - \alpha_i^{(1)}\beta_i^{(1)} \phi_{ik}^{(1)} \Cov{t}{\E{x}{E_i}}{\E{x}{F_k}} \right) \right] \nonumber \\
   & - \frac{1}{S-1} \sum\limits_{r \neq i}^S \frac{a_i}{a_r} \left[ \sum \limits_{k = 1}^L \left( \zeta_{r} \phi_{rk}^{(1)}  \Cov{x,t}{E_r}{F_k} - \alpha_r^{(1)}\beta_r^{(1)} \phi_{rk}^{(1)} \Cov{t}{\E{x}{E_r}}{\E{x}{F_k}} \right) \right]
\end{flalign} 

\tcbsubtitle{Fitness-density covariance}
      
\begin{flalign} 
    \Delta\kappa_i = & \left[ \sum \limits_{k = 1}^L \E{t}{\Cov{x}{\nu_i}{ \alpha_i^{(1)} E_i + \beta_i^{(1)} \phi_{ik}^{(1)} F_k}} \right] \nonumber \\
    & - \frac{1}{S-1} \sum\limits_{r \neq i}^S \frac{a_i}{a_r} \left[ \sum \limits_{k = 1}^L\E{t}{\Cov{x}{\nu_r}{ \alpha_r^{(1)} E_r + \beta_r^{(1)} \phi_{rk}^{(1)} F_k}} \right]
\end{flalign}     
\end{tcolorbox}

From the linear structure of the equation above, it should be clear how to further partition the coexistence mechanisms into contributions from single regulating factors, or contributions from subsets of regulating factors. We will demonstrate how this partitoning would work, using examples with a single regulating factor, $F_k$. 

\paragraph{The contribution of regulating factor $F_k$ to the linear density-dependent effects}, i.e., species $i$'s degree of specialization on $F_k$:
\begin{equation}
\begin{aligned}
\beta_i^{(1)} \phi_{ik}^{(1)} \E{x,t}{F_k - F_k^{*}}  - \frac{1}{S-1} \sum\limits_{r \neq i}^S  \frac{a_i}{a_r} \beta_r^{(1)} \phi_{rk}^{(1)} \E{x,t}{F_k - F_k^{*}} \nonumber
\end{aligned}
\end{equation}

\paragraph{The contribution of regulating factor $F_k$ to the storage effect}:

\begin{flalign}
   & \left[ \left( \zeta_{i} \phi_{ik}^{(1)}  \Cov{x,t}{E_i}{F_k} - \alpha_i^{(1)}\beta_i^{(1)} \phi_{ik}^{(1)} \Cov{t}{\E{x}{E_i}}{\E{x}{F_k}} \right) \right] \nonumber \\
   & - \frac{1}{S-1} \sum\limits_{r \neq i}^S \frac{a_i}{a_r} \left[ \left( \zeta_{r} \phi_{rk}^{(1)}  \Cov{x,t}{E_r}{F_k} - \alpha_r^{(1)}\beta_r^{(1)} \phi_{rk}^{(1)} \Cov{t}{\E{x}{E_r}}{\E{x}{F_k}} \right) \right] \nonumber
\end{flalign} 

It should also be straightforward to derive the space-time decompositions of the small-noise coexistence mechanisms. For example, the contribution of $F_k$ to the time component of the storage effect is 

\begin{flalign}
\left(\zeta_i \phi_{ik}^{(1)} - \alpha_i^{(1)}\beta_i^{(1)} \phi_{ik}^{(1)} \right) \Cov{t}{\E{x}{E_i}}{\E{x}{F_k}} 
   & - \frac{1}{S-1} \sum\limits_{r \neq i}^S \frac{a_i}{a_r}  \left(\zeta_r \phi_{rk}^{(1)} - \alpha_r^{(1)}\beta_r^{(1)} \phi_{rk}^{(1)} \right) \Cov{t}{\E{x}{E_r}}{\E{x}{F_k}}. 
\end{flalign}  

\paragraph{The contribution of regulating factor $F_k$ to relative nonlinearity}   
Arguably, there are several ways to partition relative nonlinearity further with respect to individual regulating factors.

\begin{enumerate}
\item The contribution of $F_k$'s variance to relative nonlinearity, or equivalently, the degree of specialization on the variance in $F_k$:

\begin{equation}
\begin{aligned}
& \frac{1}{2}  \left[ \left(\beta_i^{(2)} \phi_{ikk}^{(2)} \Var{x,t}{F_k} - {\beta_i^{(1)}}^2 \phi_{ik}^{(1)} \phi_{ik}^{(1)} \Var{t}{\E{x}{F_k}} \right) \right] \\
& - \frac{1}{2(S-1)} \sum\limits_{r \neq i}^S \frac{a_i}{a_r} \left[  \left(\beta_i^{(2)} \phi_{rkk}^{(2)} \Var{x,t}{F_k} - {\beta_i^{(1)}}^2 \phi_{rk}^{(1)} \phi_{rk}^{(1)} \Var{t}{\E{x}{F_k}} \right) \right]
\end{aligned}
\end{equation}

\item The contribution of covariance between $F_k$ and $F_l$ ($k \neq l$) on relative nonlinearity, or equivalently, the degree of specialization on the covariance between $F_k$ and $F_l$: 

\begin{equation}
\begin{aligned}
& \left[\left(\beta_i^{(2)} \phi_{ikl}^{(2)} \Cov{x,t}{F_k}{F_l} - {\beta_i^{(1)}}^2 \phi_{ik}^{(1)} \phi_{il}^{(1)} \Cov{t}{\E{x}{F_k}}{\E{x}{F_l}} \right) \right] \\
& - \frac{1}{S-1} \sum\limits_{r \neq i}^S \frac{a_i}{a_r} \left[  \left(\beta_i^{(2)} \phi_{rkl}^{(2)} \Cov{x,t}{F_k}{F_l} - {\beta_i^{(1)}}^2 \phi_{rk}^{(1)} \phi_{rl}^{(1)} \Cov{t}{\E{x}{F_k}}{\E{x}{F_l}} \right) \right]
\end{aligned}
\end{equation}

\item The total contribution of $F_k$ to relative nonlinearity, or equivalently, the degree of specialization on fluctuations in $F_k$, given that there are fluctuations in other regulating factors:  

\begin{equation}
\begin{aligned}
& \frac{1}{2}  \left[ \left(\beta_i^{(2)} \phi_{ikk}^{(2)} \Var{x,t}{F_k} - {\beta_i^{(1)}}^2 \phi_{ik}^{(1)} \phi_{ik}^{(1)} \Var{t}{\E{x}{F_k}} \right) \right] \\
& - \frac{1}{2(S-1)} \sum\limits_{r \neq i}^S \frac{a_i}{a_r} \left[  \left(\beta_i^{(2)} \phi_{rkk}^{(2)} \Var{x,t}{F_k} - {\beta_i^{(1)}}^2 \phi_{rk}^{(1)} \phi_{rk}^{(1)} \Var{t}{\E{x}{F_k}} \right) \right] \\
& + \left[ \sum \limits_{l \neq k}^L \left(\beta_i^{(2)} \phi_{ikl}^{(2)} \Cov{x,t}{F_k}{F_l} - {\beta_i^{(1)}}^2 \phi_{ik}^{(1)} \phi_{il}^{(1)} \Cov{t}{\E{x}{F_k}}{\E{x}{F_l}} \right) \right]  \\
& - \frac{1}{S-1} \sum\limits_{r \neq i}^S \frac{a_i}{a_r} \left[ \sum \limits_{l \neq k }^L \left(\beta_i^{(2)} \phi_{rkl}^{(2)} \Cov{x,t}{F_k}{F_l} - {\beta_i^{(1)}}^2 \phi_{rk}^{(1)} \phi_{rl}^{(1)} \Cov{t}{\E{x}{F_k}}{\E{x}{F_l}} \right) \right]
\end{aligned}
\end{equation}

\end{enumerate}

\paragraph{Exact coexistence mechanisms}

The exact coexistence mechanisms can be obtained by following the directions implied by the formulas for the exact coexistence mechanisms (\eqref{dEe} - \eqref{dkappae}). For example, the formula for $\Delta {\rho_{i}}^{(e)}$ (\eqref{drhoe}) directs the user to set $C_j$ to $\E{x,t}{C_j}$; because $C_j = \phi_j(\boldsymbol{F})$, one would set $\phi_j(\boldsymbol{F})$ to $\E{x,t}{\phi_j(\boldsymbol{F})}$.

However, this approach does not give the user much latitude to partition the coexistence mechanisms further into contributions from individual regulating factors or subsets of regulating factors. Returning to the task of partitioning $\Delta {\rho_{i}}^{(e)}$, we might try setting $\phi_j(F_1, \ldots, F_L)$ to $\phi_j(F_1, \ldots, \E{x,t}{F_k}, \ldots, F_l)$, one $F_k$ at a time, and then summing the resulting pieces of invasion growth rate to approximate $\Delta {\rho_{i}}^{(e)}$. This approach will make for a good approximation of $\Delta {\rho_{i}}^{(e)}$ under the small-noise assumptions. Similar procedures can be defined for $\Delta N_{i}^{(e)}$ and $\Delta I_{i}^{(e)}$.

To define these procedures in a reasonable amount of page-space, new notation is required. Let $\boldsymbol{F}^{-\{k\}}$ be the vector of regulating factors with the $k$-th element removed. A natural extension is $\boldsymbol{v}^{-\{k,l\}}$, a vector $\boldsymbol{v}$ where the $k$-th and $l$-th elements have been removed. Let $\left\{\boldsymbol{v}^{-\{k\}}, a \right\}$ be a vector $\boldsymbol{v}$ where the $k$-th element has been replaced with $a$. Similarly, let $\left\{\boldsymbol{v}^{-\{k,l\}}, a, b \right\}$ be a vector $\boldsymbol{v}$ where the $k$-th element has been replaced with $a$, and the $l$-th element has been replaced by $b$. The notation introduced here allows us to express ideas such as holding all elements of $\boldsymbol{F}$ at their equilibrium values, except for $F_k$, which is held at its spatiotemporal average: $\left\{\boldsymbol{F^*}^{-\{k\}}, \E{x,t}{F_k} \right\}$ 

\begin{tcolorbox}[breakable, title = Formulas for exact coexistence mechanisms: multiple regulating factors, subtitle style={boxrule=0.4pt,
colback=black!30!white}]

\tcbsubtitle{The invasion growth rate}

\begin{equation} \label{exact_co_mech}
    \E{t}{\log(\widetilde{\lambda}_{i})} = {\Delta E_{i}}^{'(e)} + {\Delta\rho_i}^{'(e)} + {\Delta\mathrm{N}_i}^{'(e)} + {\Delta\mathrm{I}_i}^{'(e)} + {\Delta\kappa_i}^{'(e)},
\end{equation}

\tcbsubtitle{Density-independent effects}

\begin{flalign} \label{r'e}
     {\Delta E_{i}}^{'(e)} & = \E{t}{\log(\E{x}{g_i(E_i,\phi_i(\boldsymbol{F^{*i}})})}  - \frac{1}{S-1} \sum\limits_{r \neq i}^S \frac{a_i}{a_r} \E{t}{\log(\E{x}{g_r(E_r,\phi_r(\boldsymbol{F^{*r}})})} 
\end{flalign}

\tcbsubtitle{Linear density-dependent effects}

\begin{flalign} \label{drhoe}
 {\Delta \rho_i}^{'(e)} & = \left[ \sum \limits_{k = 1}^{L} \log( g_i \left( E_i^*, \phi_i \left( \left\{ \boldsymbol{F^{*i}}^{-\{k\}}, \E{x,t}{F_k} \right\} \right) \right)) \right]\\
 & - \frac{1}{S-1} \sum\limits_{r \neq i}^S \frac{a_i}{a_r} \left[ \sum \limits_{k = 1}^{L}  \log( g_r \left( E_r^*, \phi_r \left( \left\{ \boldsymbol{F^{*r}}^{-\{k\}}, \E{x,t}{F_k} \right\} \right) \right)) \right]    
\end{flalign}

\tcbsubtitle{Relative nonlinearity}

\begin{flalign} 
 {\Delta N_i}^{'(e)} & =  \Biggl[ \sum \limits_{k = 1}^{L} \sum \limits_{l = 1}^{L} \E{t}{\log( \E{x}{g_i \left( E_i^*, \phi_i \left( \left\{\boldsymbol{F^{*i}}^{-\{k,l\}}, F_k, F_l \right\} \right) \right)})}  \\
 & - \frac{1}{S-1} \sum\limits_{r \neq i}^S \frac{a_i}{a_r} \Biggl[ \sum \limits_{k = 1}^{L} \sum \limits_{l = 1}^{L} \E{t}{\log( \E{x}{g_r \left( E_r^*, \phi_r \left( \left\{\boldsymbol{F^{*r}}^{-\{k,l\}}, F_k, F_l \right\} \right) \right)})} \Biggr] \\
 & = \Biggl[ \sum \limits_{k = 1}^{L} \sum \limits_{l = 1}^{L} \E{t}{\log( \E{x}{g_i \left( E_i^*, \phi_i \left( \left\{\boldsymbol{F^{*i}}^{-\{k,l\}}, F_k, F_l \right\} \right) \right)})} \\
 & -  \log( g_i \left( E_i^*, \phi_i \left( \left\{\boldsymbol{F^{*i}}^{-\{k\}}, \E{x,t}{F_k} \right\} \right) \right))  \\
 & -  \log( g_i \left( E_i^*, \phi_i \left( \left\{\boldsymbol{F^{*i}}^{-\{l\}}, \E{x,t}{F_l} \right\} \right) \right)) \Biggr] \\
 & - \frac{1}{S-1} \sum\limits_{r \neq i}^S \frac{a_i}{a_r} \Biggl[ \sum \limits_{k = 1}^{L} \sum \limits_{l = 1}^{L} \E{t}{\log( \E{x}{g_r \left( E_r^*, \phi_r \left( \left\{\boldsymbol{F^{*r}}^{-\{k,l\}}, F_k, F_l \right\} \right) \right)})} \\
 & -  \log( g_r \left( E_r^*, \phi_r \left( \left\{\boldsymbol{F^{*r}}^{-\{k\}}, \E{x,t}{F_k} \right\} \right) \right)) \\
 & -  \log( g_r \left( E_r^*, \phi_r \left( \left\{\boldsymbol{F^{*r}}^{-\{l\}}, \E{x,t}{F_l} \right\} \right) \right)) \Biggr]
\end{flalign}      
      
\tcbsubtitle{The storage effect}

\begin{align*} 
 {\Delta I_i}^{'(e)} & = \Biggl[ \sum \limits_{k = 1}^{L} \E{t}{\log( \E{x}{g_i \left( E_i, \phi_i \left( \left\{\boldsymbol{F^{*i}}^{-\{k\}}, F_k \right\} \right) \right)})} - \overline{\mathscr{E}_i}' \\
  & \quad \quad - \sum \limits_{l = 1}^{L} \E{t}{\log( \E{x}{g_i \left( E_i^*, \phi_i \left( \left\{\boldsymbol{F^{*i}}^{-\{k,l\}}, F_k, F_l \right\} \right) \right)})} \Biggr] \\
 & - \frac{1}{S-1} \sum\limits_{r \neq i}^S \frac{a_i}{a_r} \Biggl[ \sum \limits_{k = 1}^{L} \E{t}{\log( \E{x}{g_r \left( E_r, \phi_r \left( \left\{\boldsymbol{F^{*r}}^{-\{k\}}, F_k\right\} \right) \right)})}  - \overline{\mathscr{E}_r}' \\
  & \quad - \sum \limits_{l = 1}^{L} \E{t}{\log( \E{x}{g_r \left( E_r^*, \phi_r \left( \left\{\boldsymbol{F^{*r}}^{-\{k,l\}}, F_k, F_l \right\} \right) \right)})} \Biggr]
\end{align*}      

\tcbsubtitle{Fitness-density covariance}

\begin{flalign} \label{dkappae'}
    \Delta {\kappa_{i}}^{'(e)} =  \E{t}{\log(\widetilde{\lambda}_{i})} - \left({\Delta E_{i}}^{'(e)} + {\Delta\rho_i}^{'(e)} + {\Delta\mathrm{N}_i}^{'(e)} + {\Delta\mathrm{I}_i}^{'(e)}\right)
\end{flalign}  

\end{tcolorbox}

The exact coexistence mechanisms here are slightly different from the exact coexistence mechanisms presented in the main text (\eqref{dEe} - \eqref{dkappae}), hence the "prime" notation. This is because the exact coexistence mechanisms here are deliberately defined in terms of sums across regulating factors; this property enables more fine-grained partitioning into the effects of particular regulating factors, or various space-time decompositions. Under the small-noise assumptions, the two types of exact coexistence mechanisms do not differ by more than $\mathcal{O}\sigma^3$. We note that different partitions may serve different purposes

The linear density-dependent effects is obtained by fixing the environment and most of the regulating factors at their equilibrium values, with the exception that a single regulating factor is fixed at its spatiotemporal average: $\E{x,t}{F_k}$. The growth rate is calculated under these conditions, and then procedure is repeated for each regulating factor. The sum of results is ${\Delta \rho_i}^{'(e)}$. 

Relative nonlinearity is obtained by fixing the environment and most regulating factors at their equilibrium levels, except now two regulating factors are allowed to vary. Since the growth rate calculated under these conditions includes the average effects of regulating factors, we must subtract the average effect (e.g., $\log( g_j \left( E_j^*, \phi_j \left( \left\{\boldsymbol{F^{*j}}^{-\{l\}}, \E{x,t}{F_l} \right\} \right) \right))$) of each of the regulating factors that are varying. Corresponding growth rates are summed across all pairs of regulating factors to produce ${\Delta N_{i}}^{'(e)}$. 

The storage effect is obtained by allowing the environment and a single regulating factor to vary. In this scenario, the emergent growth rate includes the main effect of the environment, the spatiotemporal average effect of the regulating factor, and effects of fluctuations in pairs of regulating factors. Therefore, in order to isolate the interaction between the environment and a regulating factor, we must subtract $\overline{\mathscr{E}_i}'$. We must also subtract the appropriate terms within  ${\Delta \rho_i}^{'(e)}$ and ${\Delta N_{i}}^{'(e)}$, which amounts to $\sum_{l = 1}^{L} \E{t}{\log( \E{x}{g_r \left( E_r^*, \phi_r \left( \left\{\boldsymbol{F^{*r}}^{-\{k,l\}}, F_k, F_l \right\} \right) \right)})}$.

\subsubsection{Structured population models}
\label{app:Generalization of MCT to different classes of models:Structured population models}

In a structured population model, discrete age, stage, or size classes can be treated as different regulating factors $F_k$, and thus handled via the methods above (Section \ref{app:Generalization of MCT to different classes of models:Continuous-time models:Multiple regulating factors}). When population structure is continuous in nature, the methods above can still be applied, but instead of holding some regulating factors at their equilibrium values while allowing others to vary, one would hold a range of a single (continuous) regulating factor constant, while allowing another range to vary. In other words, sums across discrete regulating factors are replaced by an integral across the states of a single continuous regulating factor.

\subsection{The maximum number of species that can coexist via fitness density covariance }
\label{app:The maximum number of species that can coexist via fitness density covariance }

When there is no temporal variation, \eqref{fd_cov_ec} in Appendix \ref{app:Deriving the small-noise fitness-density covariance for the spatiotemporal lottery model} reduces to
\begin{equation}
\begin{aligned}
\E{t}{\Cov{x}{v_j}{\lambda_j}} & \approx \frac{q}{1-q} \Var{x}{\alpha_j^{(1)} (E_j(x) - E_{j}^{*}) + \beta_j^{(1)} (C_r(x) - C_{r}^{*})} \\
 & \approx \frac{q}{1-q} \left[ {\alpha_j^{(1)}}^2 \Var{x}{E_j} + 2 \alpha_j^{(1)} \beta_j^{(1)} \Cov{x}{E_j}{C_j} + {\beta_j^{(1)}}^2 \Var{x}{C_j} \right]. 
\end{aligned}
\end{equation}
The competition parameter $C$ can be expanded as a function of $L$ regulating factors (see Appendix \ref{app:Generalization of MCT to different classes of models:Continuous-time models:Multiple regulating factors}):$C_j = \phi_j(\boldsymbol{F})$, where $\boldsymbol{F} = (F_1, F_2, ..., F_L)$. The environment parameter can be expressed a vector of $M$ discrete states: $E_j \in \boldsymbol{E}'$, where $\boldsymbol{E}' = (E'_{1}, E'_{2}, ..., E'_{M})$. Then, there are $M \times L$ "effective regulating factors" with the form $\Cov{x}{E'_{m}}{F_k}$; and $L(L-1)/2$ "effective regulating factors" with the form $\Cov{x}{F_k}{F_l}$. This result shows that fitness density covariance can potentially support a large number of species. 

\section{References}
\printbibliography[heading=none]

\end{document}